\documentclass[]{pasj01}

\input{pasjsup.sty}

\newcounter{author}
\def\authorcount#1#2{\refstepcounter{author}\label{#1}
\altaffiltext{\ref{#1}}{#2}}

\begin{document}
\SetRunningHead{Y. Wakamatsu et al.}{Title}

\Received{201X/XX/XX}
\Accepted{201X/XX/XX}

\title{ASASSN-18aan: An Eclipsing SU UMa-type Cataclysmic Variable 
with a 3.6-hour Orbital Period and a Late G-type Secondary Star}

\author{Yasuyuki~\textsc{Wakamatsu},\altaffilmark{\ref{affil:AA}*}
John R.~\textsc{Thorstensen},\altaffilmark{\ref{affil:AB}}
Naoto~\textsc{Kojiguchi},\altaffilmark{\ref{affil:AA}}
Keisuke~\textsc{Isogai},\altaffilmark{\ref{affil:AA},\ref{affil:AC}}
Mariko~\textsc{Kimura},\altaffilmark{\ref{affil:AA},\ref{affil:AD}}
Ryuhei~\textsc{Ohnishi},\altaffilmark{\ref{affil:AA}}
Taichi~\textsc{Kato},\altaffilmark{\ref{affil:AA}}
Hiroshi~\textsc{Itoh},\altaffilmark{\ref{affil:AE}}
Yuki~\textsc{Sugiura},\altaffilmark{\ref{affil:AF}}
Sho~\textsc{Sumiya},\altaffilmark{\ref{affil:AF}}
Hanami~\textsc{Matsumoto},\altaffilmark{\ref{affil:AF}}
Daiki~\textsc{Ito},\altaffilmark{\ref{affil:AF}}
Kengo~\textsc{Nikai},\altaffilmark{\ref{affil:AF}}
Hiroshi~\textsc{Akitaya},\altaffilmark{\ref{affil:AG},\ref{affil:AH}}
Chihiro~\textsc{Ishioka},\altaffilmark{\ref{affil:AI}}
Kohei~\textsc{Oide},\altaffilmark{\ref{affil:AI}}
Takahiro~\textsc{Kanai},\altaffilmark{\ref{affil:AI}}
Yoshinori~\textsc{Uzawa},\altaffilmark{\ref{affil:AJ}}
Yumiko~\textsc{Oasa},\altaffilmark{\ref{affil:AG},\ref{affil:AI},\ref{affil:AJ}}
Tam\'as~\textsc{Tordai},\altaffilmark{\ref{affil:AK}}
Tonny~\textsc{Vanmunster},\altaffilmark{\ref{affil:AL}}
Sergey Yu.~\textsc{Shugarov},\altaffilmark{\ref{affil:AM},\ref{affil:AN}}
Masayuki~\textsc{Yamanaka},\altaffilmark{\ref{affil:AC},\ref{affil:AH}}
Mahito~\textsc{Sasada},\altaffilmark{\ref{affil:AH}}
Kengo~\textsc{Takagi},\altaffilmark{\ref{affil:AH}}
Yuki~\textsc{Nishinaka},\altaffilmark{\ref{affil:AH}}
Yuina~\textsc{Yamazaki},\altaffilmark{\ref{affil:AH}}
Ikki~\textsc{Otsubo},\altaffilmark{\ref{affil:AH}}
Tatsuya~\textsc{Nakaoka},\altaffilmark{\ref{affil:AH}}
Katsuhiro L.~\textsc{Murata},\altaffilmark{\ref{affil:AO}}
Ryou~\textsc{Ohsawa},\altaffilmark{\ref{affil:AP},\ref{affil:AQ}}
Masahiro~\textsc{Morita},\altaffilmark{\ref{affil:AP}}
Makoto~\textsc{Ichiki},\altaffilmark{\ref{affil:AP}}
Sjoerd~\textsc{Dufoer},\altaffilmark{\ref{affil:AR}}
Masanori~\textsc{Mizutani},\altaffilmark{\ref{affil:AS}}
Takashi~\textsc{Horiuchi},\altaffilmark{\ref{affil:AT}}
Miyako~\textsc{Tozuka},\altaffilmark{\ref{affil:AU}}
Masaki~\textsc{Takayama},\altaffilmark{\ref{affil:AU}}
Tomohito~\textsc{Ohshima},\altaffilmark{\ref{affil:AU}}
Tomoki~\textsc{Saito},\altaffilmark{\ref{affil:AU}}
Pavol A.~\textsc{Dubovsky},\altaffilmark{\ref{affil:AV}}
Geoff~\textsc{Stone},\altaffilmark{\ref{affil:AW}}
Ian~\textsc{Miller},\altaffilmark{\ref{affil:AX}} and
Daisaku~\textsc{Nogami}\altaffilmark{\ref{affil:AA}}
}

\authorcount{affil:AA}{
Department of Astronomy, Kyoto University, Kyoto 606-8502, Japan}
\authorcount{affil:AB}{
Department of Physics and Astronomy, 6127 Wilder Laboratory, Dartmouth College, Hanover, NH 03755-3528, USA}
\authorcount{affil:AC}{
Okayama Observatory, Kyoto University, 3037-5 Honjo, Kamogatacho, Asakuchi, Okayama 719-0232, Japan}
\authorcount{affil:AD}{
Extreme Natural Phenomena RIKEN Hakubi Research Team, Cluster for Pioneering Research, RIKEN, 2-1 Hirosawa, Wako, Saitama 351-0198, Japan}
\authorcount{affil:AE}{
Variable Star Observers League in Japan (VSOLJ), 1001-105 Nishiterakata, Hachioji, Tokyo 192-0153, Japan}
\authorcount{affil:AF}{
Osaka Kyoiku University, 4-698-1 Asahigaoka, Kashiwara, Osaka 582-8582, Japan}
\authorcount{affil:AG}{
Graduate School of Science and Engineering, Saitama University, 255 Shimo-Okubo, Sakura-ku, Saitama, Saitama 338-8570, Japan}
\authorcount{affil:AH}{
Hiroshima Astrophysical Science Center, Hiroshima University, 1-3-1 Kagamiyama, Higashi-Hiroshima, Hiroshima 739-8526, Japan}
\authorcount{affil:AI}{
Graduate School of Education, Saitama University, 255 Shimo-Okubo, Sakura-ku, Saitama, Saitama 338-8570, Japan}
\authorcount{affil:AJ}{
Faculty of Education, Saitama University, 255 Shimo-Okubo, Sakura-ku, Saitama, Saitama 338-8570, Japan}
\authorcount{affil:AK}{
Polaris Observatory, Hungarian Astronomical Association, Laborc utca 2/c, 1037 Budapest, Hungary}
\authorcount{affil:AL}{
Center for Backyard Astrophysics Belgium, Walhostraat 1A, B-3401 Landen, Belgium}
\authorcount{affil:AM}{
Sternberg Astronomical Institute, Lomonosov Moscow State University, Universitetsky Ave., 13, Moscow 119992, Russia}
\authorcount{affil:AN}{
Astronomical Institute of the Slovak Academy of Sciences, 05960 Tatranska Lomnica, Slovakia}
\authorcount{affil:AO}{
Department of Physics, Tokyo Institute of Technology, 2-12-1 Ookayama, Meguro-ku, Tokyo 152-8551, Japan}
\authorcount{affil:AP}{
Institute of Astronomy, Graduate School of Science, The University of Tokyo, 2-21-1 Osawa, Mitaka, Tokyo 181-0015, Japan}
\authorcount{affil:AQ}{
Kiso Observatory, Institute of Astronomy, Graduate School of Science, The University of Tokyo 10762-30, Mitake, Kiso-machi, Kiso-gun, Nagano 397-0101, Japan}
\authorcount{affil:AR}{
Vereniging Voor Sterrenkunde (VVS), Oostmeers 122 C, 8000 Brugge, Belgium}
\authorcount{affil:AS}{
Variable Star Observers League in Japan (VSOLJ), Okayama, Japan}
\authorcount{affil:AT}{
Ishigakijima Astronomical Observatory, Public Relations Center, National Astronomical Observatory of Japan, 1024-1 Arakawa, Ishigaki, Okinawa, 907-0024, Japan}
\authorcount{affil:AU}{
Nishi-Harima Astronomical Observatory, Center for Astronomy, University of Hyogo, 407-2, Nishigaichi, Sayo-cho, Sayo, Hyogo 679-5313, Japan}
\authorcount{affil:AV}{
Vihorlat Observatory, Mierova 4, 06601 Humenne, Slovakia}
\authorcount{affil:AW}{
American Association of Variable Star Observers, 49 Bay State Rd., Cambridge, MA 02138, USA}
\authorcount{affil:AX}{
Furzehill House, Ilston, Swansea, SA2 7LE, UK}

\email{$^*$wakamatsu@kusastro.kyoto-u.ac.jp}


\maketitle

\begin{abstract}
We report photometric and spectroscopic observations of the eclipsing
SU UMa-type dwarf nova ASASSN-18aan. We observed the 2018
superoutburst with 2.3 mag brightening and found the orbital period
($P_{\rm orb}$) to be 0.149454(3) d, or 3.59 hr.  This is longward of
the period gap, establishing ASASSN-18aan as one of a small number
of long-$P_{\rm orb}$ SU UMa-type dwarf novae. The estimated
mass ratio, ($q=M_2/M_1 = 0.278(1)$), is almost identical to the
upper limit of tidal instability by the 3:1 resonance.
From eclipses, we found that the accretion disk at the onset of the
superoutburst may reach the 3:1 resonance radius, suggesting that
the superoutburst of ASASSN-18aan results from the tidal instability.
Considering the case of long-$P_{\rm orb}$ WZ Sge-type dwarf novae,
we suggest that the tidal dissipation at the tidal truncation radius is
enough to induce SU UMa-like behavior in relatively high-$q$ systems
such as SU UMa-type dwarf novae, but that this is no longer effective in
low-$q$ systems such as WZ Sge-type dwarf novae.
The unusual nature of the system extends to the secondary star,
for which we find a spectral type of G9, much earlier than 
typical for the orbital period, and a secondary mass
$M_2$ of around 0.18 M$_{\odot}$, smaller than expected
for the orbital period and the secondary's spectral type.
We also see indications of enhanced sodium
abundance in the secondary's spectrum.  Anomalously 
hot secondaries are seen in a modest number of other CVs 
and related objects. These systems evidently
underwent significant nuclear evolution before the onset 
of mass transfer. In the case of ASASSN-18aan, this apparently
resulted in a mass ratio lower than typically found at
the system's $P_{\rm orb}$, which may account for the occurrence
of a superoutburst at this relatively long period.

\end{abstract}

\section{Introduction}
\label{sec:intro}
Cataclysmic variables (CVs) are close binary systems consisting of
a white dwarf primary star and a secondary star
that fills its Roche lobe and transfers mass to the primary
via the inner Lagrangian point $L_1$.  Unless the white dwarf
is highly magnetized, an accretion disk forms around the
primary.  Dwarf novae (DNe) are a subclass of CVs 
characterized by recurrent outbursts of 2-10 magnitude,
evidently caused by accretion disk instabilities.

Over most of the life of a CV, the binary separation gradually decreases
due to angular momentum
loss from the system, resulting shortening of the orbital period. The mass
ratio also decreases due to the mass transfer from the secondary to the
primary. CVs therefore generally evolve toward extreme states with low
mass ratios and short orbital periods. Relatively few non-magnetic 
CVs are observed in a period range from roughly 2 to 3 hr,
known as the period gap
(see \cite{kni11CVdonor} and references therein for more 
on CV evolution).

SU UMa-type DNe, a subclass of DNe, show long-lasting, brighter outbursts
called {\it superoutbursts} in addition to (normal) outbursts.
It is a key feature of superoutbursts to be accompanied by superhumps,
which are variations of small amplitude, typically 0.1-0.5 mag, and have 
slightly longer periods than orbital periods. 
The superoutburst is considered to be a result of the tidal instability that is
triggered when the outer disk reaches the 3:1 resonance radius
(\cite{whi88tidal}; \cite{hir90SHexcess}; \cite{lub91SHa}; \cite{lub91SHa}).
The expansion of the accretion disk to the 3:1 resonance radius can only
occur if the mass ratio $q = M_2/M_1$ is less than 0.25, \citep{whi88tidal}, 
or 0.33 with reduction of the mass transfer from the secondary
\citep{mur00SHintermediateq}. The period of most SU UMa-type DNe lie
below the period gap; at these short periods, the mass rations are typically
$< 0.25$ The expansion of the accretion disk is limited by the tidal truncation
radius, where the angular momentum of orbiting material is removed via tidal
torque. The 3:1 resonance, therefore, is considered to occur when the tidal
truncation radius is larger than the 3:1 resonance radius. The detailed
action of the tidal truncation radius, however, is still unknown and it is unclear
whether it acts as a hard limit of disk expansion, especially in extremely
low-$q$ systems \citep{osa02wzsgehump}.

A small number of CVs with longer periods have shown the superhumps
and superoutbursts characteristic of SU UMa stars; these 
long-$P_{\rm orb}$ objects should offer tests of the tidal instability model
and the effects of tidal torques at the tidal truncation radius.

The longest-known of these long-period SU UMa stars is TU Men
\citep{sto84tumen}, which has an orbital period $P_{\rm orb}=0.1172(2)$,
or 2.81 hr \citep{men95tumen}. The mass ratio of TU Men is estimated
by various authors; 0.58 \citep{sto84tumen},
0.455(45) or 0.33 \citep{men95tumen}, 0.5(2) \citep{sma06tumen}.
\citet{sma06tumen} also presented the lower limit of $q>0.41(8)$.
TU Men has been treated as an exceptional SU UMa-type DN because of its
long orbital period and anomalously large mass ratio, which is far
beyond the upper critical value of the 3:1 resonance predicted by the tidal
instability model.

Other SU UMa-type DNe have been detected in the period gap 
(see, e.g., \citet{kat03CVperiodgap}, \citet{pav14nyser}).
More recently, some SU UMa-type DNe have been discovered above the period gap.
In 2018, CS Ind \citep{kat19csind} showed the superoutburst accompanied by
superhumps with the mean superhump period $P_{\rm}=0.12471(1)$ d
and the behavior of the superoutburst is similar to that of WZ Sge-type
DNe, which is a subclass of SU UMa-type DNe and have extremely short
orbital periods and small mass ratios (see \citet{kat15wzsge} for a review).
OT J002656.6+284933 showed superoutburst in 2016 and has an unlikely
small mass ratio, less than 0.15, in spite of its long superhump period of
0.13225(1) d \citep{kat17j0026}. ASASSN-14ho is another curious SU UMa-type
DNe. This object showed four rebrightenings after the main superoutburst in 2014
\citep{kat20asas14ho}. The orbital period and mass ratio are 0.24315(10) d and
0.28, respectively \citep{Gas19}. This orbital period is the longest one among known
SU UMa-type DNe. Additionally, the secondary of ASASSN-14ho is anomalously
warm for its orbital period; its spectral type is K4 $\pm$ 2 although the expected type
from the standard model \citep{kni06CVsecondary} is M0 \citep{Gas19}.
These CVs with anomalously hot secondaries are considered to be formed when the
secondary underwent significant nuclear evolution prior to the onset of the mass
transfer \citep{tho15asassn13cl}.

It should be noted that \citet{sma04ugemSH} reported the detection of superhumps
in the 1985 long outburst of U Gem. The orbital period and mass ratio of U Gem are
0.1769061911(28) \citep{mar90ugem} and 0.364(17) \citep{sma01ugem},
respectively. If U Gem indeed underwent a superoutburst, this presents a serious
challenge to the tidal instability model since the mass ratio of U Gem is too large to
grow tidal instability.

In addition to long-$P_{\rm orb}$ SU UMa-type DNe, some
WZ Sge-type DNe have been reported that have atypically long orbital periods.
ASASSN-16eg \citep{wak17asassn16eg}, which showed superoutburst
accompanied by clear early superhumps, has long orbital period and
anomalously large mass ratio, almost twice the upper limit of the 2:1
resonance. \citet{osa02wzsgehump} suggested that in extremely low
$q$-systems such as WZ Sge-type DNe, the tidal torque
by the secondary at the tidal truncation radius is weak and the outer disk could
reach the 2:1 resonance radius beyond the tidal truncation radius.

In this paper, we report observations of ASASSN-18aan 
($\alpha = 0^{\rm h}\ 46^{\rm m}\ 08^{\rm s}.05, 
\delta = +62^{\circ}\ 10'\ 04''.9$, from the \cite{GaiaPaper2}). 
The superoutburst was detected on 2018 November 30 by the All-Sky Automated Survey for
Supernovae (ASAS-SN: \cite{ASASSN}). 
The inverse of the Gaia DR2 parallax is 675 (+32, $-$29) pc
\citep{GaiaPaper2}.
\citet{gre18} computed three-dimensional reddening
maps across the northern sky\footnote{One can query
these maps at http://argonaut.skymaps.info/}.  Their
map gives $E(g-r) = 0.35$ at the Gaia distance along
this line of sight; using expressions found in 
\citet{gre19} we convert this to $A_V = 1.19$ mag,
or $E(B - V) = 0.36$ mag, where we have
assumed $R = A_V / E(B-V) = 3.32$.
\citet{nes19a18aan} obtained an optical spectrum indicating the spectral
type of the secondary is G--K. They also investigated historical
outbursts of ASASSN-18aan from plates and found that the recurrence
time may be about 11 months. The object showed superhump-like
modulations in the 2018 superoutbursts, as well as clear eclipses. 
The eclipses revealed on orbital period that was quite long for an
SU UMa-type DN, 0.1495 d as a tentative value, which attracted special interest
(vsnet-alert 22806, 22810). Observations showed the
superhump period to be much longer than the orbital one, indicating an
unusually large mass ratio (vsnet-alert 22816, 22817, 22821). We
therefore performed world-wide photometric and spectroscopic
observations. We also performed photometric and spectroscopic
observations in quiescent state to clarify its nature and binary
parameters. Our observations are described in section 2, and their
results are in section 3. We describe the analysis in section4, and
discuss the nature of this anomalous object in section 5. We
summarize our conclusions in section 6.

\begin{table*}
\caption{Summary of OISTER observations.}
\centering
\begin{tabular}{p{6.0cm}ccccc}
\hline \hline
Telescope (Instrument) & Date (UT) & Exposure (Sec) & \multicolumn{3}{c}{Filter (Number of data)} \\
\hline
SaCRA (MuSaSHI) 0.55 m & 2018-12-10 & 30 & \multicolumn{3}{c}{r (810), i (810), z (810)} \\
& 2018-12-14 & 60 & \multicolumn{3}{c}{r (206), i (206), z (206)} \\
& 2018-12-17 & 60 & \multicolumn{3}{c}{r (258), i (258), z (258)} \\
& 2018-12-19 & 60 & \multicolumn{3}{c}{r (488), i (488), z (488)} \\
& 2018-12-23 & 60 & \multicolumn{3}{c}{r (248), i (248), z (248)} \\
& 2019-01-07 & 60 & \multicolumn{3}{c}{r (115), i (115), z (115)}  \\
\hline
Kiso (Tomo-e Gozen) 1.05 m & 2018-12-10 & 7.5 & \multicolumn{3}{c}{No filter (1181)}  \\
\hline
MITSuME Akeno (Tricolor camera) 0.5 m & 2018-12-10 & 60 & \multicolumn{3}{c}{g (79), Rc (79), Ic (79)} \\
& 2018-12-12 & 60 & \multicolumn{3}{c}{g (105), Rc (105), Ic (105)} \\
& 2018-12-13 & 60 & \multicolumn{3}{c}{g (25), Rc (25), Ic (25)} \\
& 2018-12-14 & 60 & \multicolumn{3}{c}{g (106), Rc (106), Ic (106)} \\
& 2018-12-15 & 60 & \multicolumn{3}{c}{g (106), Rc (106), Ic (106)} \\
\hline
MITSuME Okayama (Tricolor camera) 0.5 m & 2018-12-08 & 60 & \multicolumn{3}{c}{g (52), Rc (52), Ic (52)} \\
\hline
Kanata (HOWPol) 1.5 m & 2018-12-10 & 40 & \multicolumn{3}{c}{I (318)} \\
& 2018-12-13 & 40 & \multicolumn{3}{c}{I (74)} \\
& 2018-12-15 & 40 & \multicolumn{3}{c}{I (141)} \\
& 2018-12-18 & 40 & \multicolumn{3}{c}{I (108)} \\
& 2018-12-19 & 40 & \multicolumn{3}{c}{I (74)} \\
& 2018-12-27 & 60 & \multicolumn{3}{c}{V (82), I (69)} \\
& 2019-01-07 & 60 & \multicolumn{3}{c}{V (42), I (53)} \\
& 2019-01-13 & 60 & \multicolumn{3}{c}{V (65), I (66)} \\
& 2019-01-14 & 60 & \multicolumn{3}{c}{V (37), I (75)} \\
& 2019-01-15 & 60 & \multicolumn{3}{c}{V (32), I (41)} \\
& 2019-01-16 & 60 & \multicolumn{3}{c}{V (27), I (41)} \\
& 2019-02-02 & 60 & \multicolumn{3}{c}{V (62), I (61)} \\
\hline
Murikabushi (Tricolor camera) 1.05 m & 2018-12-20 & 60 & \multicolumn{3}{c}{g (218), Rc (218), Ic (218)} \\
& 2018-12-24 & 60 & \multicolumn{3}{c}{g (97), Rc (97), Ic (97)} \\
\hline
\hline
Telescope (Instrument) & Date (UT) & Exposure (Sec) & Cover range & Resolution & Number of data \\
\hline
Nayuta (MALLS) 2.0 m & 2018-12-10 & 1200 & 5130-7970 & 1800 & 16 \\
& 2018-12-24 &1200 & 4490-9650 & 800 & 3  \\
& 2019-01-05 & 1200 & 4390-9550 & 800 & 3 \\
\hline
\hline
\end{tabular}
\label{tab:OISTER}
\end{table*}

\begin{table*}
\caption{Journal of MDM Time-Series Photometry}
\begin{tabular}{ccccc}
\hline \hline
Start (UT) & HA range & Exp. (Sec) & $N$ & Filter \\
\hline
2019-09-05 07:38  &  $-$01:36 -- $+$02:41  &  30  &  510  &  GG420  \\
2019-09-06 07:11  &  $-$01:59 -- $+$02:15  &  60  &  249  &  R  \\
2019-09-08 06:15  &  $-$02:48 -- $+$01:40  &  30  &  357  &  GG420  \\
2019-10-22 03:37  &  $-$02:30 -- $+$02:43  &  30  &  553  &  GG420  \\
2019-10-23 07:18  &  $+$01:16 -- $+$03:42  &  30  &  274  &  GG420  \\
2019-10-24 03:03  &  $-$02:56 -- $+$02:11  &  33  &  475  &  GG420  \\
\hline
\hline
\end{tabular}
\label{tab:MDMphot}
\end{table*}

\section{Observations}
\subsection{Spectroscopy}
Our most extensive set of spectra is from the 
Ohio State Multi-Object Spectrograph (OSMOS; \cite{mar11}) mounted 
on the 2.4 m Hiltner telescope at MDM Observatory,
on Kitt Peak, Arizona.  We obtained a sequence of eighteen 720 s 
exposures of ASASSN-18aan on 2019 Sept. 06 UT.  
The spectra span 4.12 hr from start to finish, 
and cover just over one orbit.  The spectra cover from
3975 to 6890 \AA , with 0.7 \AA\ pixel$^{-1}$ and
3.0 \AA\ resolution (full width at half-maximum).
We derived a pixel-to-wavelength relation from 
Hg, Ne, and Xe lamp spectra, and used 
the [OI] and OH-band features in the night
sky spectrum to derive zero point offsets
and linear stretch factors that we applied
to the wavelength scales of the individual exposures.  
To convert to absolute flux, we observed spectrophotometric
standard stars in twilight when the sky was reasonably 
clear.  

\subsection{MDM Time Series Photometry}

Our MDM time series photometry (Table \ref{tab:MDMphot}) is all from the McGraw-Hill
1.3m telescope.  In 2019 September we used an 
Andor frame-transfer camera on three nights; exposures
were generally 30 seconds with almost no dead time
between exposures.
In 2019 October we used a SITe 1024$^2$ CCD detector
cropped to 256$^2$, which resulted in a $\sim 3$ s 
dead time.  For most of the observations we
used a Schott GG420 glass filter, which blocks 
$\lambda < 4200$ \AA\ -- functionally similar
to no filter -- but for one night in September we used an
$r$ filter.  The same set of comparison stars was used
througout.  The main comparison star, at
$\alpha = 0^{\rm h} 46^{\rm m} 05^{\rm s}.8, \delta = +62^{\circ} 10'
43''.4$ (J2000), is listed at $V = 14.446$ in 
APASS \citep{hen11aavso}.  We adjusted our differential magnitudes
by this amount to convert them to rough $V$ magnitudes.

\subsection{VSNET \& OISTER Time Series Photometry and spectroscopy}
\label{sec:obs_so}
We performed a world-wide observational campaign via Variable
Star Network (VSNET) collaborations \citep{VSNET} and Optical
and Infrared Synergetic Telescopes for Education and Research
(OISTER). All of our photometric observations by VSNET and OISTER were
described in barycentric Julian date (BJD). Detailed logs of
photometric observations by VSNET and OISTER are listed in table
E2. Our intensive observations of the
superoutburst of ASASSN-18aan by VSNET was started on
December 4, 2018 (BJD 2458457), 4 days after the ASAS-SN's detection.

We also performed spectroscopic observations via OISTER to confirm
changes of spectral features through the superoutburst on 2018-12-10,
2018-12-24 and 2019-01-05 (UT). We could not observed standard stars
due to a bad weather and time limitation so not performed a sensitivity
correction.

Table \ref{tab:OISTER} gives a summary of the OISTER observations.

\section{Results}
\subsection{Overall light curve of superoutburst}
\label{sec:lightcurve}
\begin{figure*}
\begin{center}
\FigureFile(120mm,150mm){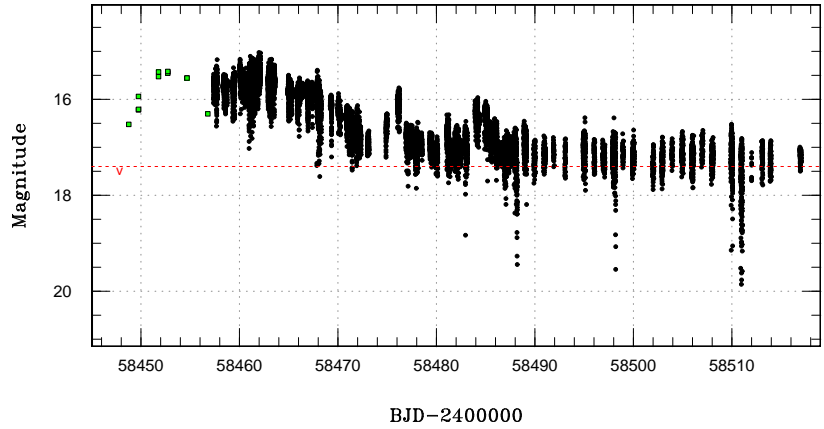}
\end{center}
\caption{Light curve of ASASSN-18aan. The filled- squared points show the ASAS-SN's V magnitude. The V-shaped sign represents an upper limit by ASAS-SN's V and g magnitude, and the quiescent magnitude is shown as a red dashed line.}
\label{fig:a18aan_lightcurve}
\end{figure*}

The overall light curve of the superoutburst is shown in figure
\ref{fig:a18aan_lightcurve}. We also plot photometric data from the
ASAS-SN CV Patrol (\cite{ASASSN}; \cite{Koc17ASASSN};
\cite{dav15ASASSNCVAAS}) that constrains the onset of the superoutburst.
The superoutburst lasted about 21 days during BJD 2458452-2458473.
The brightness first rapidly increased during BJD 2458448-2458452 and
next slightly increased during BJD 2458452-2458462, reaching maximum
at 15.2 mag, and then began to decrease. The main superoutburst
ended at BJD 2458473. There were two rebrightenings (during
BJD 2458474-2458477 and 2458483-2458486) after the main superoutburst.
After the rebrightenings, the magnitude returned to near the quiescent state
around 17.5 mag. The superoutburst amplitude was $\sim$ 2.3 mag.

\subsection{Orbital period}
For analysis, we first subtracted the global trend of the light curve of the
superoutburst and rebrightenings by subtracting a smoothed light curve
obtained by locally weighted polynomial regression (LOWESS: \cite{LOWESS})
After the subtraction, we performed a phase dispersion minimization (herafter
PDM, \cite{PDM}) analysis of whole superoutburst and found the orbital period to
be $P_{\rm orb}=0.149454(3)$ d. We then derived the
phase-averaged profile of the eclipse folded on the orbital period. To determine
the mid-eclipse time, we fitted the Gaussian profile and derived BJD
2458461.428617 as the epoch of the eclipse.

\subsection{Superhump period}
\label{sec:sh}
We removed all eclipses to investigate the variation of the superhump period
during the superoutburst. We calculated the time of eclipses from the above
epoch and orbital period and masked eclipses with a range of 0.11 in units of
orbital phase.

\begin{figure}
\begin{center}
\FigureFile(80mm,100mm){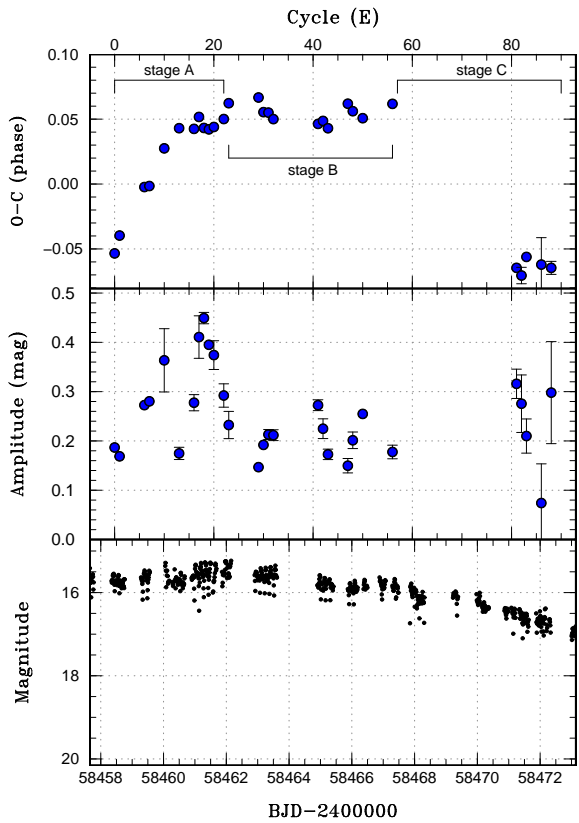}
\end{center}
\caption{Upper panel: $O-C$ curve of ASASSN-18aan during
BJD 2458458-58472. We used an ephemeris of BJD 2458461.485+0.1579$E$ for
drawing this figure. Middle panel: Amplitude of superhumps. Lower panel:
Light curve. The horizontal axis in units of BJD and cycle number is common to
all of these panels.}
\label{fig:a18aan_ocamp}
\end{figure}

Figure \ref{fig:a18aan_ocamp} shows the $O-C$ curve (upper panel), the
amplitude of the superhumps (middle panel), and the light curve (lower panel)
of ASASSN-18aan during BJD 2458458-58474. We determined the times of
maxima of superhumps in the same way as in \citet{Pdot}. The resulting times
are listed in table E3.

From the variation of the superhump period and the amplitudes of superhumps,
we regarded BJD 2458455.0-2458462.5 ($0 \leq E \leq 22$) as stage A, BJD
2458462.5-2458467.6 ($29 \leq E \leq 56$) as stage B, and BJD 2458467.6-2458473.5
($60 \leq E \leq 93$) as stage C superhumps.

\begin{figure}
\begin{center}
\FigureFile(70mm,100mm){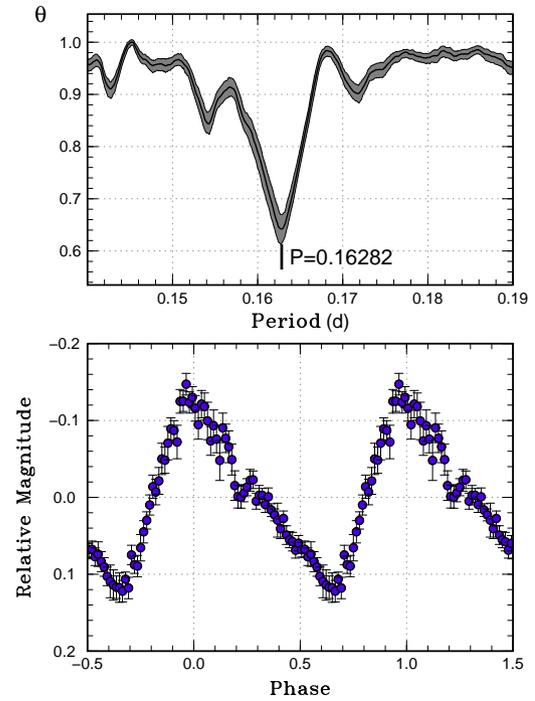}
\end{center}
\caption{Upper panel: $\theta$-diagram of our PDM analysis of stage A
superhumps of ASASSN-18aan (BJD 2458455.0-2458462.5). The gray area
represents the 1$\sigma$ error of the best-estimated period by the PDM method.
Lower panel: Phase-averaged profile of stage A superhumps.}
\label{fig:a18aan_pdmA}
\end{figure}

The PDM analysis of stage A superhumps (upper panel) and the phase-averaged
mean profile (lower panel) are shown in figure \ref{fig:a18aan_pdmA}. 
Single-peaked, asymmetric modulations are evident, with an amplitude of 0.3 mag.
We found the stage A superhump period to be $P_{\rm stA} = 0.16282(5)$ d. We
also found the stage B and stage C superhump period to be 0.15821(4) d and
0.14944(5) d, respectively. Note that stage C superhump period is equal to the orbital
period within the error. This indicates that ellipsoidal variations by the secondary 
contaminated superhumps in stage C. To derive a plausible period of stage C
superhumps, we subtracted the phase-averaged profile of quiescent light curve
in BJD 2458490-2458520, which shows apparent ellipsoidal variations. After
the subtraction, we performed PDM analysis again and derived the period of
0.15743(6) d. This period is shorter than that of stage B by 0.5 \% and consistent
with ordinary cases \citep{Pdot}{; therefore, we adopt} this as the stage
C superhump period.

$P_{\rm dot} (\equiv \dot{P}_{\rm sh}/P_{\rm sh})$, which is the derivative of the
superhump period during stage B, is $91(34) \times 10^{-5}$. This value is quite
large, but less reliable because of the large error and uncertainty of the boundary
of each superhump stage stemmed from small number of points.

\subsection{Variation of phase-averaged profile through superoutburst}
\label{sec:phave}
\begin{figure}
\begin{center}
\FigureFile(70mm,100mm){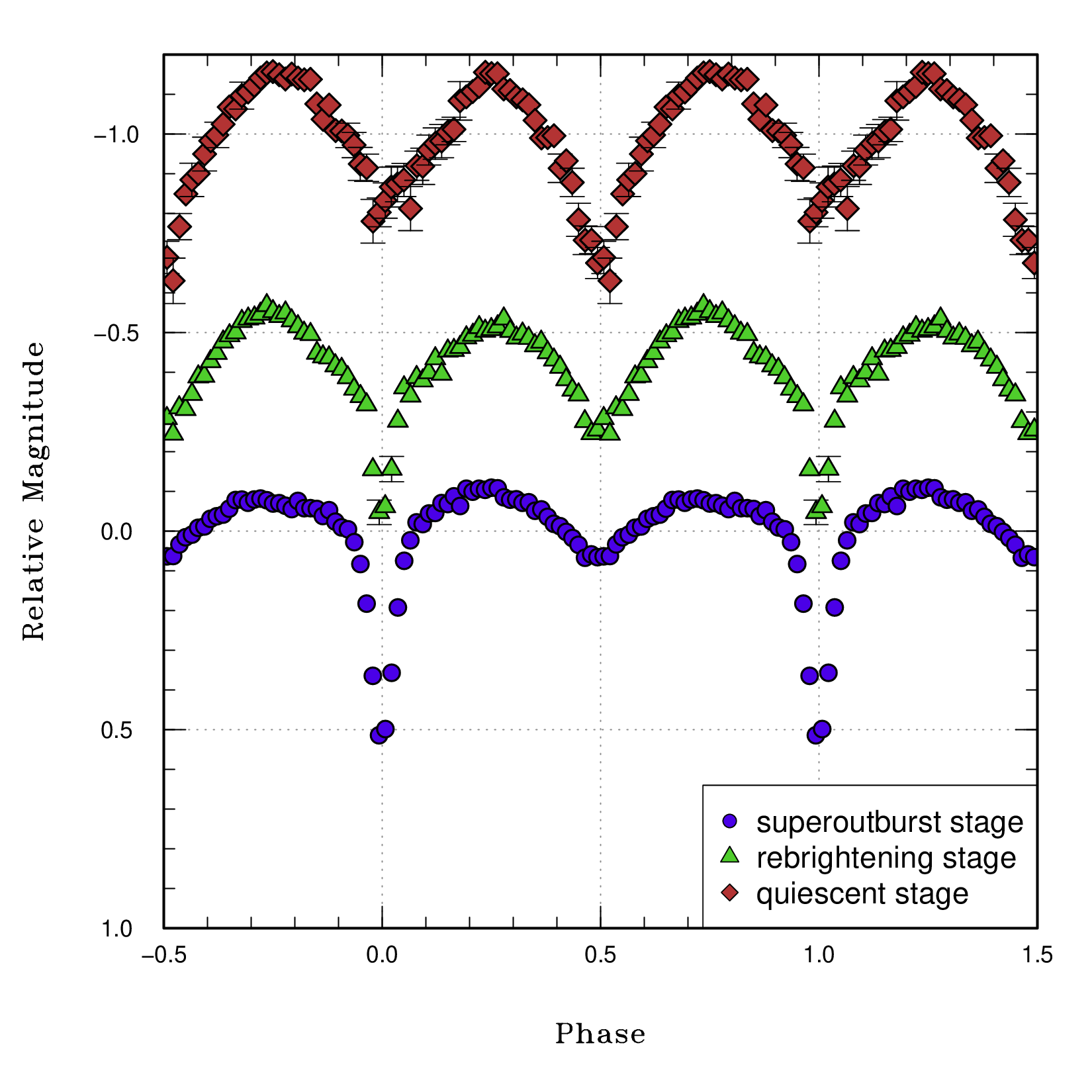}
\end{center}
\caption{Phase-averaged profiles of ASASSN-18aan during the superoutburst stage (BJD 2458453-2458473.5; blue circles), rebrightening stage (BJD 2458473.5-2458487; green triangles) and quiescent stage (BJD 2458487-2458520; red diamonds). }
\label{fig:phave}
\end{figure}

Figure \ref{fig:phave} shows the phase-averaged profile folded on the orbital
period. We divided the light curve of the superoutburst into three stages;
superoutburst stage (BJD 2458453-2458473.5) including a whole main
superoutburst, rebrightening stage (BJD 2458473.5-2458487) including two
rebrightenings, and quiescent stage (BJD 2458487-2458520) which is after
the rebrightenings. It is evident that the eclipse clearly seen at the superoutburst
stage around phase 0.0 becomes shallow at the rebrightening stage, and then it
can be hardly seen at the quiescent stage. This indicates the inclination $i$ of this
system is small for an eclipsing system ($i \sim 80^{\circ}$), and the eclipses 
at the superoutburst and rebrightening stages are caused by the secondary 
star passing the front of the
expanded, bright disk. It is clearly seen that the light curve at the superoutburst
or rebrightening stage is affected by ellipsoidal variations. We discussed the
detail of these ellipsoidal variations based on the MDM observation data as below.

\subsection{Variation of spectral characteristics through superoutburst}
\label{sec:spectra}

\begin{figure}
\begin{center}
\FigureFile(85mm,100mm){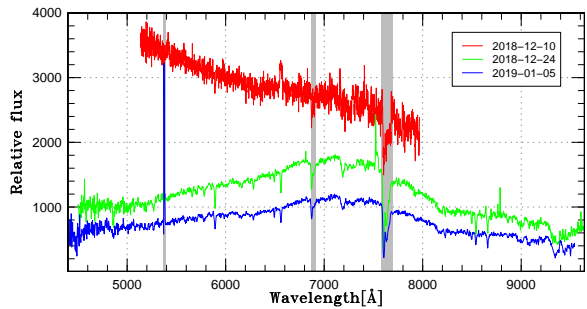}
\end{center}
\caption{Average flux. The gray area represents a contamination of the atmosphere or noise. 
{The spectra shown are from} 2018-12-10, 2018-12-24 and 2019-01-05 (UT) from top to bottom, respectively. 
The trace for 2018-12-10 is shifted above by 1600 units for convenience. The 
spectra are not flux-calibrated
(see section \ref{sec:obs_so}).}
\label{fig:spectravary}
\end{figure}

Figure \ref{fig:spectravary} shows spectra through the superoutburst. 
The spectra were obtained during the brightening state, just after 
the first rebrightening, and in the quiescent state after the second rebrightening. 
As described in section \ref{sec:obs_so}, although a sensitivity correction
was not carried out in all observations the profile transition can be seen by
comparing each trace. The disk component rose at shorter wavelength during the
superoutburst, while it became weak at the quiescent state. In addition to this,
{the H$\alpha$ $\lambda$ 6563 line went from emission in outburst to absorption 
in quiescence,} indicating an appearance of the secondary component.

\subsection{Photometric observations in quiescent state}

\begin{figure}
\begin{center}
\FigureFile(90mm,100mm){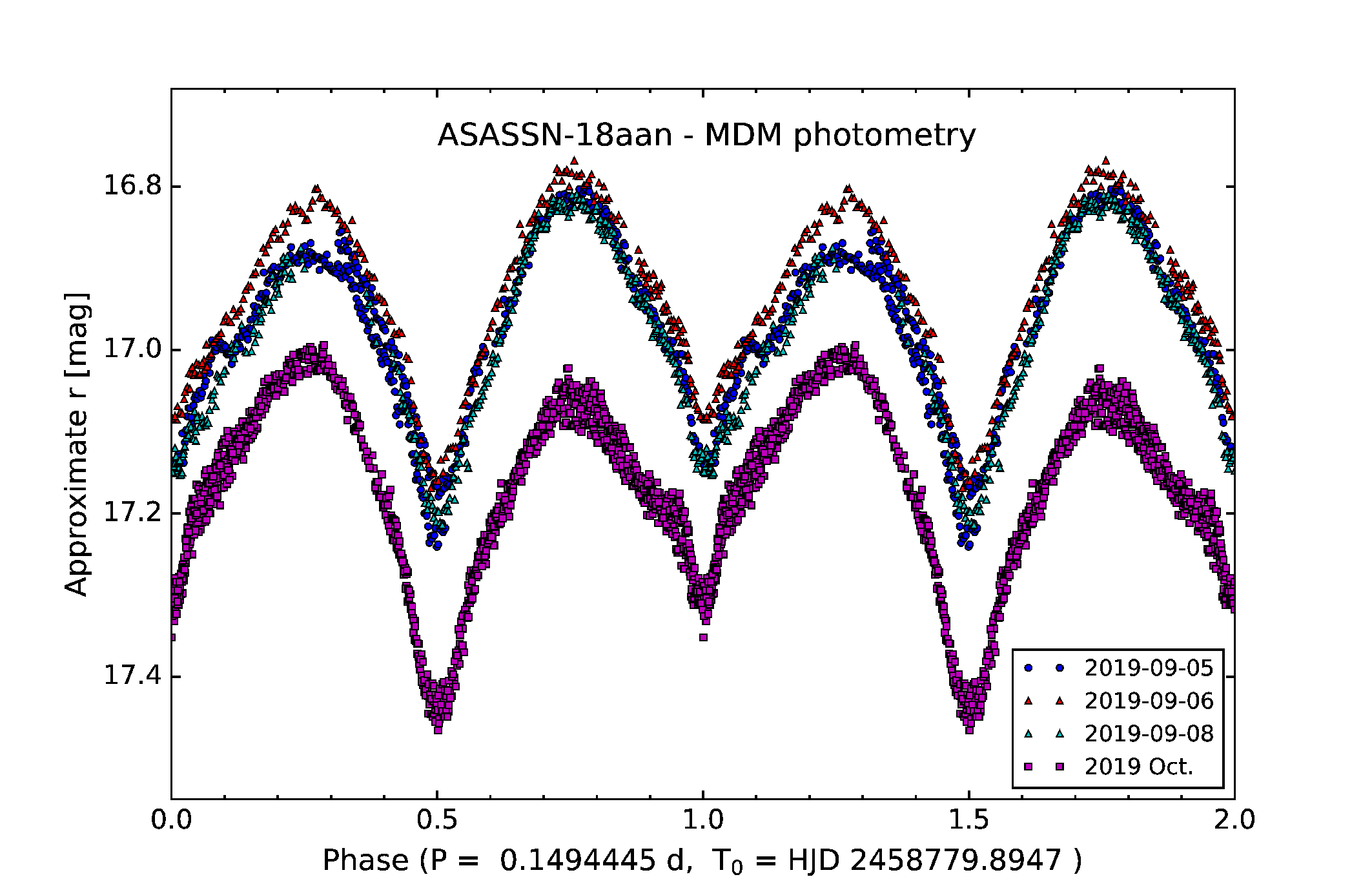}
\end{center}
\caption{Light curves from the MDM 1.3 m telescope, folded on the 
eclipse ephemeris, and repeated for a second cycle for continuity.
}
\label{fig:mdmlightcurves}
\end{figure}

Figure \ref{fig:mdmlightcurves} shows MDM light curves from 2019 September and
October, after the source had returned to quiesence.
The light curves show a persistent modulation
with two peaks per orbit.  This ellipsoidal
variation indicates a strong contribution from a 
tidally-distorted secondary star.  Such strong
ellipsoidal variation is seldom observed in CVs with 
periods this short.  

Several features of the quiescent light curve are not consistent
with a pure ellipsoidal variation.  Most obviously, shallow
eclipses appear around phase zero.  The mean magnitude is
about 0.1 mag fainter in 2019 October than in the previous
month, while the secondary star should be nearly constant;
evidently the disk faded over the interval between observations.  
In the September data, the peaks near phase 0.75 are
slightly brighter than those near 0.25, while in the 
October data the opposite is true; in pure ellipsoidal
variation they should be exactly equal.
Also, there are subtle but clear breaks in the slope near 
phases 0.4 and 0.6.  These are clearest in the October data,
where the mean light level is a bit fainter.  The interval
around phase 0.5 was covered on all three nights in the 
October data, so this feature is highly reproducible.  We
consider its interpretation later.

\begin{figure}
\begin{center}
\FigureFile(90mm,100mm){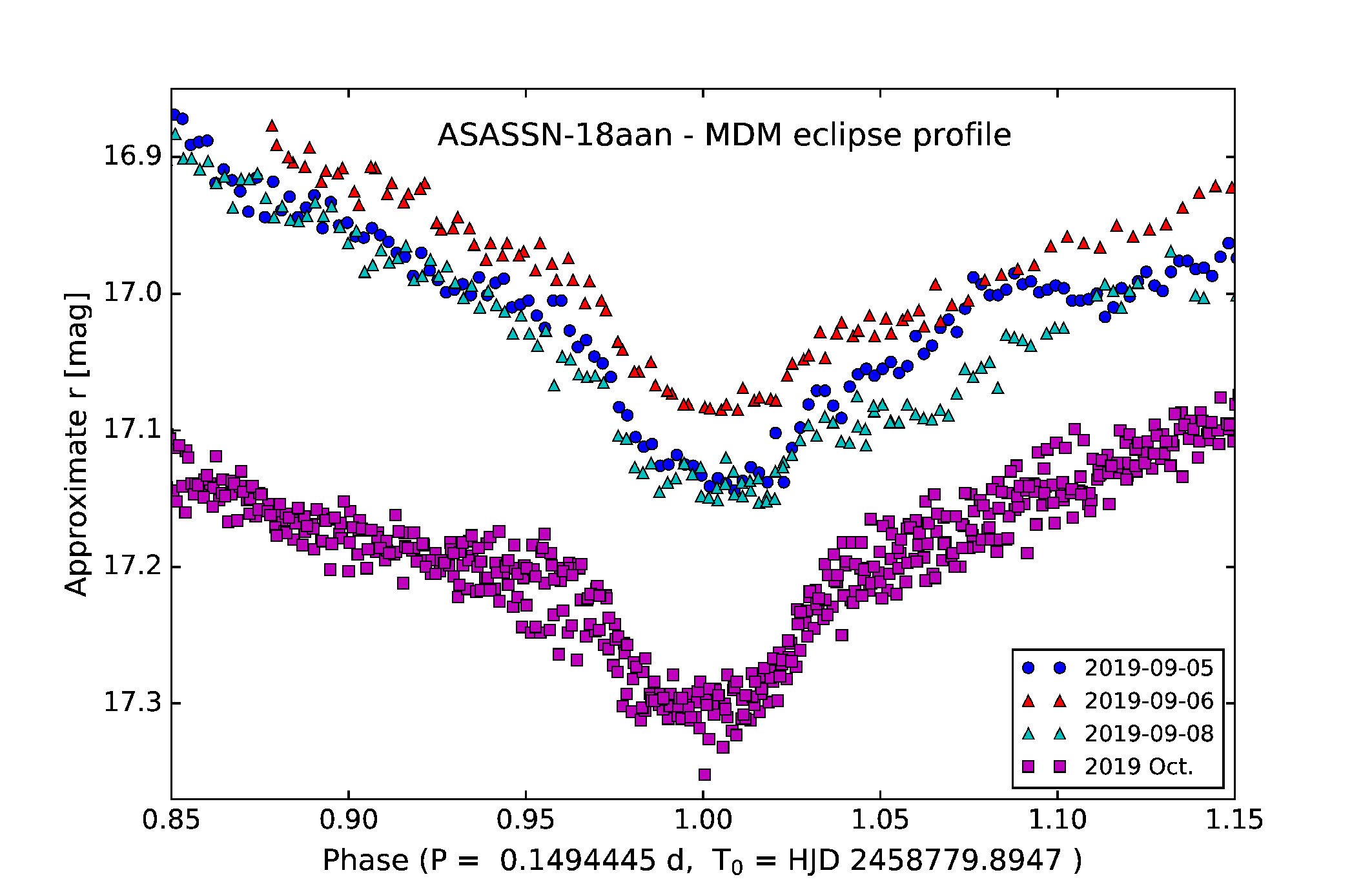}
\end{center}
\caption{Magnified light curves of the primary eclipses around phase 1.0. Other conditions are same as figure \ref{fig:mdmlightcurves}.}
\label{fig:mdmeclipse}
\end{figure}

Figure \ref{fig:mdmeclipse} gives a magnified view of the 
eclipse in the quiescent light curves.  The eclipse
is only $\sim 0.1$ mag deep, but it is clearly
detected in all the data sets, with ingress starting around
phase 0.96 and last contact around phase 0.04.  
The eclipse morphology is reminiscent of the longer-period
CV ASASSN-15aa (see Figs.~12 and 13 of \cite{tho16}),
which also shows a shallow eclipse superposed on a 
strong ellipsoidal variation.

\subsection{Spectroscopy and velocities} 

\begin{figure}
\begin{center}
\FigureFile(90mm,100mm){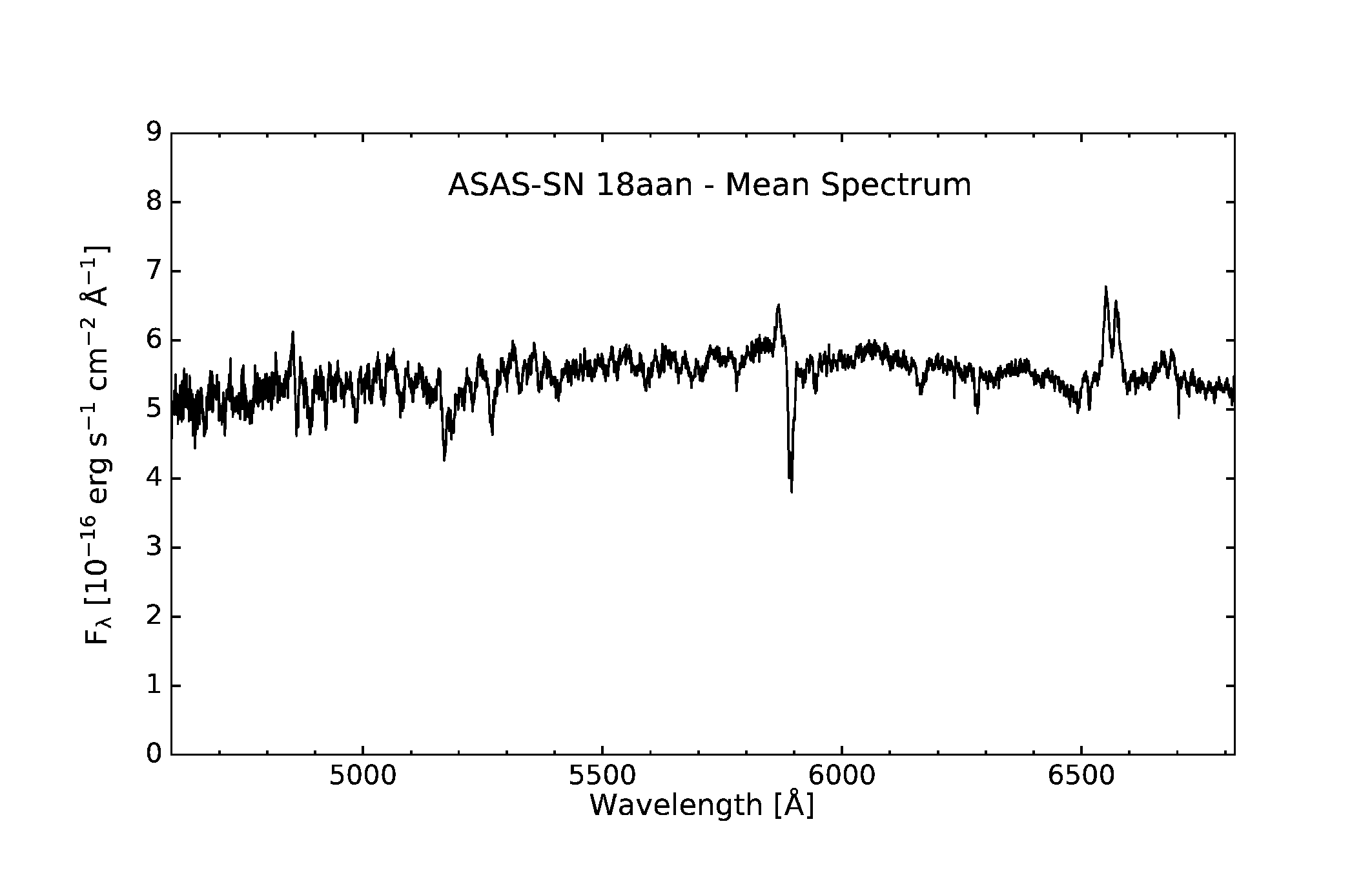}
\end{center}
\caption{Average flux-calibrated spectrum from 2019 Sept.  
}
\label{fig:specplot}
\end{figure}

The mean spectrum (figure \ref{fig:specplot}) shows weak emission, but 
also absorption features characteristic of a late-type star, as might be
expected given the strong ellipsoidal modulation.

\begin{figure}
\begin{center}
\FigureFile(90mm,100mm){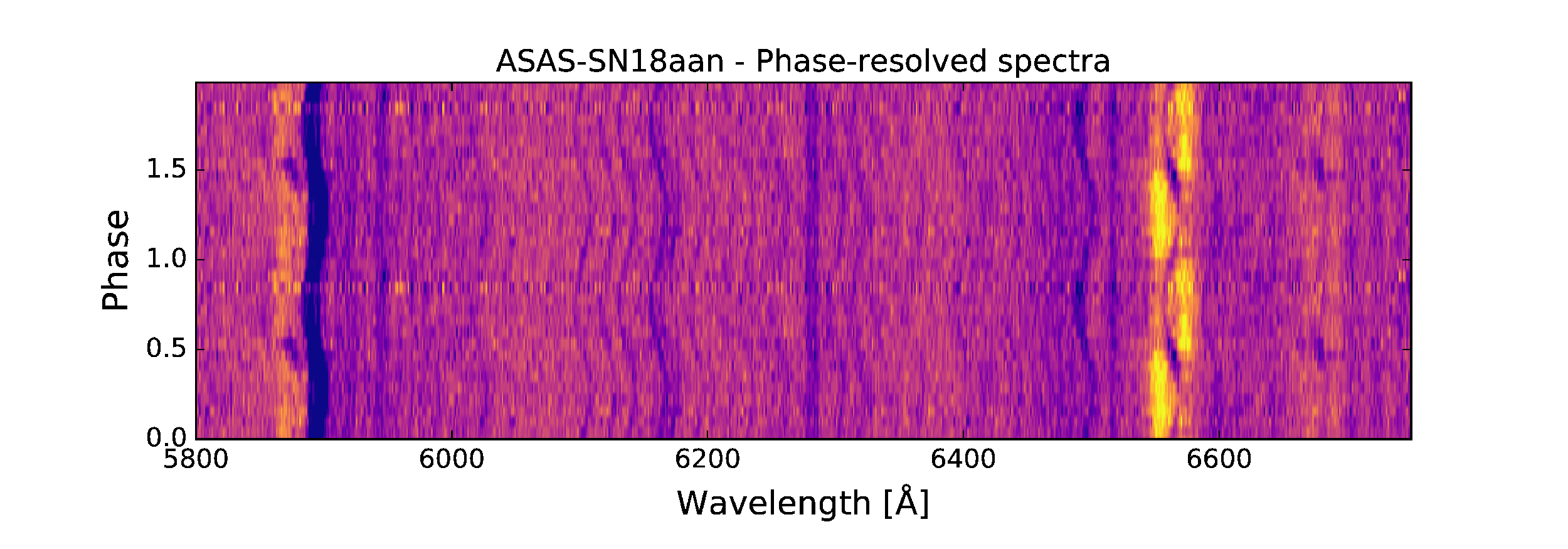}
\end{center}
\caption{Phase-resolved spectrum.  Note the radial
velocity variation of the absorption lines.}
\label{fig:singletrail}
\end{figure}

To explore how the spectrum behaves through the
orbital cycle, we prepared a phase-averaged 
representation of our spectral data.  To start,
we rectified each spectrum by dividing by 
its fitted continuum and computed the phase
of each spectrum using the photometric ephemeris.  
We then constructed a grid of equally-spaced phases,
and at each phase computed an average of the 
phase-adjacent rectified spectra, weighted by a 
narrow Gaussian centered on the fiducial phase.
Figure \ref{fig:singletrail} shows a portion of the
two-dimensional image formed by stacking these
averages.  The absorption lines vary strikingly 
in velocity on the orbital period, conclusively 
demonstrating that they are from the secondary 
star rather than some other object.  Emission 
is visible at H$\alpha$ $\lambda 6563$, but its
profile is complicated by the secondary's 
H$\alpha$ absorption line, which reveals itself
through its velocity modulation. 

Note that the emission lines at H$\alpha$ and 
\ion{He}{1} $\lambda\lambda 5876, 6678$ all
show transient absorption features
near phase 0.5.  We will return to this later.

\begin{figure}
\begin{center}
\FigureFile(90mm,100mm){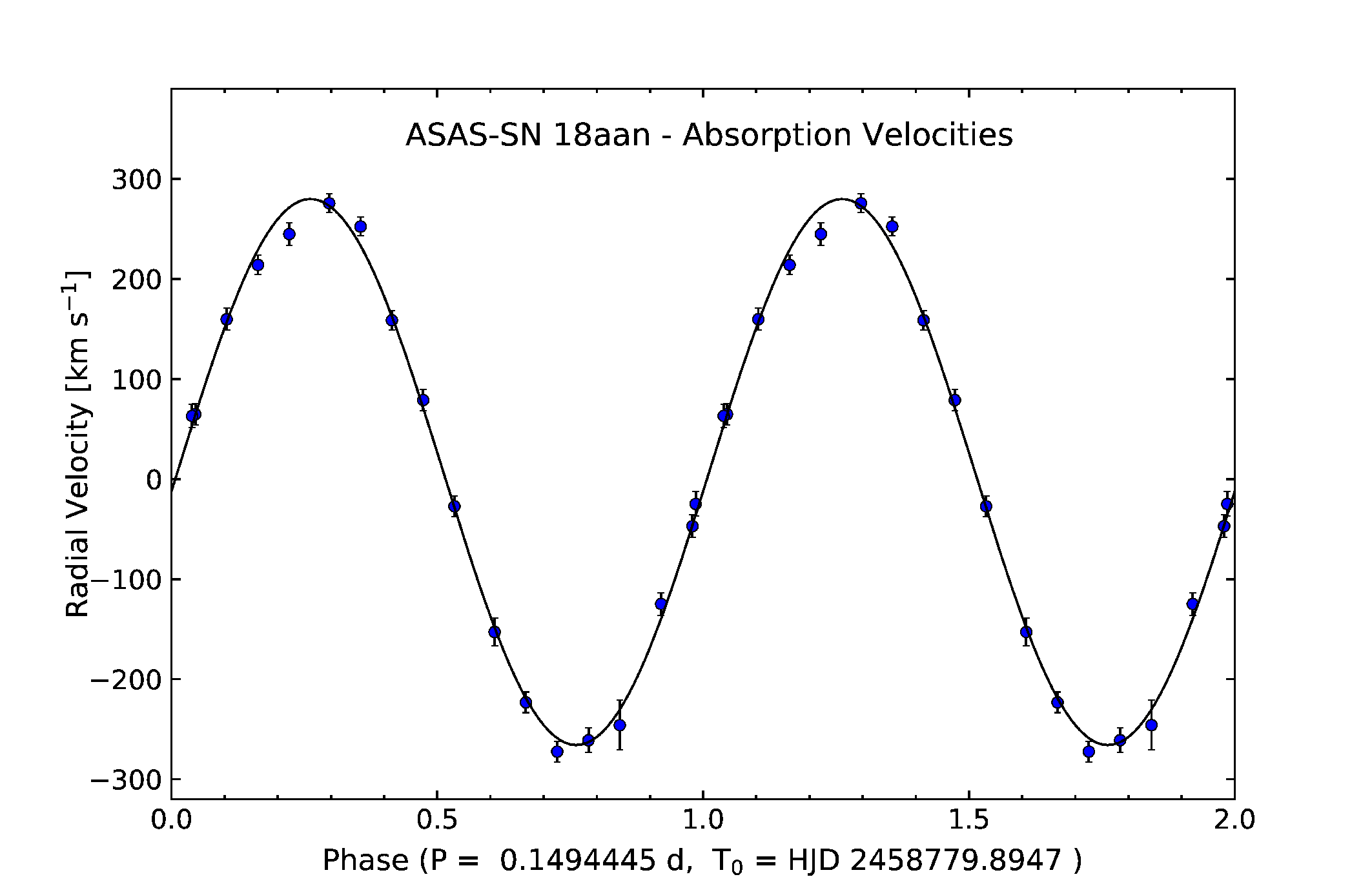}
\end{center}
\caption{Absorption-line velocities folded on the 
orbital ephemeris, with the best-fit sinusoid superposed.  All
points are repeated over a second cycle for continuity.
}
\label{fig:foldedv}
\end{figure}

We measured the radial velocities of the secondary 
component by cross-correlating the observed 
spectrum with a template spectrum 
using the IRAF task {\tt fxcor};  
\citet{ton79} describe the principle. The template
was a composite of 76 spectra of later-type 
IAU velocity standards that had been shifted to 
zero apparent radial velocity before averaging.  The
correlations excluded the region near the sodium
D lines because of possible confusion with 
\ion{He}{1} $\lambda 5876$ emission. All our 
spectra gave good correlations despite the 
exclusion of this strong feature.  We corrected
our velocities $v$ to the reference frame 
of the solar system barycenter and fitted them with a 
sinusoid of the form
\begin{equation}
v(t) = \gamma + K \sin\left[2 \pi (t - T_0) \over P\right],
\end{equation}
where $t$ is the barycentric time of mid-integration.
We fixed $P$ to the eclipse period, and found
\begin{eqnarray}
\gamma &=& 7 \pm 4 \hbox{ km s}^{-1} \\
K &=& 273 \pm 6 \hbox{ km s}^{-1}, \\
T_0 &=& \hbox{BJD } 2458732.9708 \pm 0.0005, \hbox{  and} \\
\sigma &=& 13 \hbox{ km s}^{-1}, 
\end{eqnarray}
where the time base is UTC and $\sigma$ is the standard 
deviation around the best fit.  Allowing $P$ to float
yielded $P = 0.1486(9)$ d, consistent with the much
more accurate eclipse-based period, and had almost no 
effect on the values of the other parameters. Figure \ref{fig:foldedv}
shows the radial velocities and the fit.

\begin{figure}
\begin{center}
\FigureFile(90mm,100mm){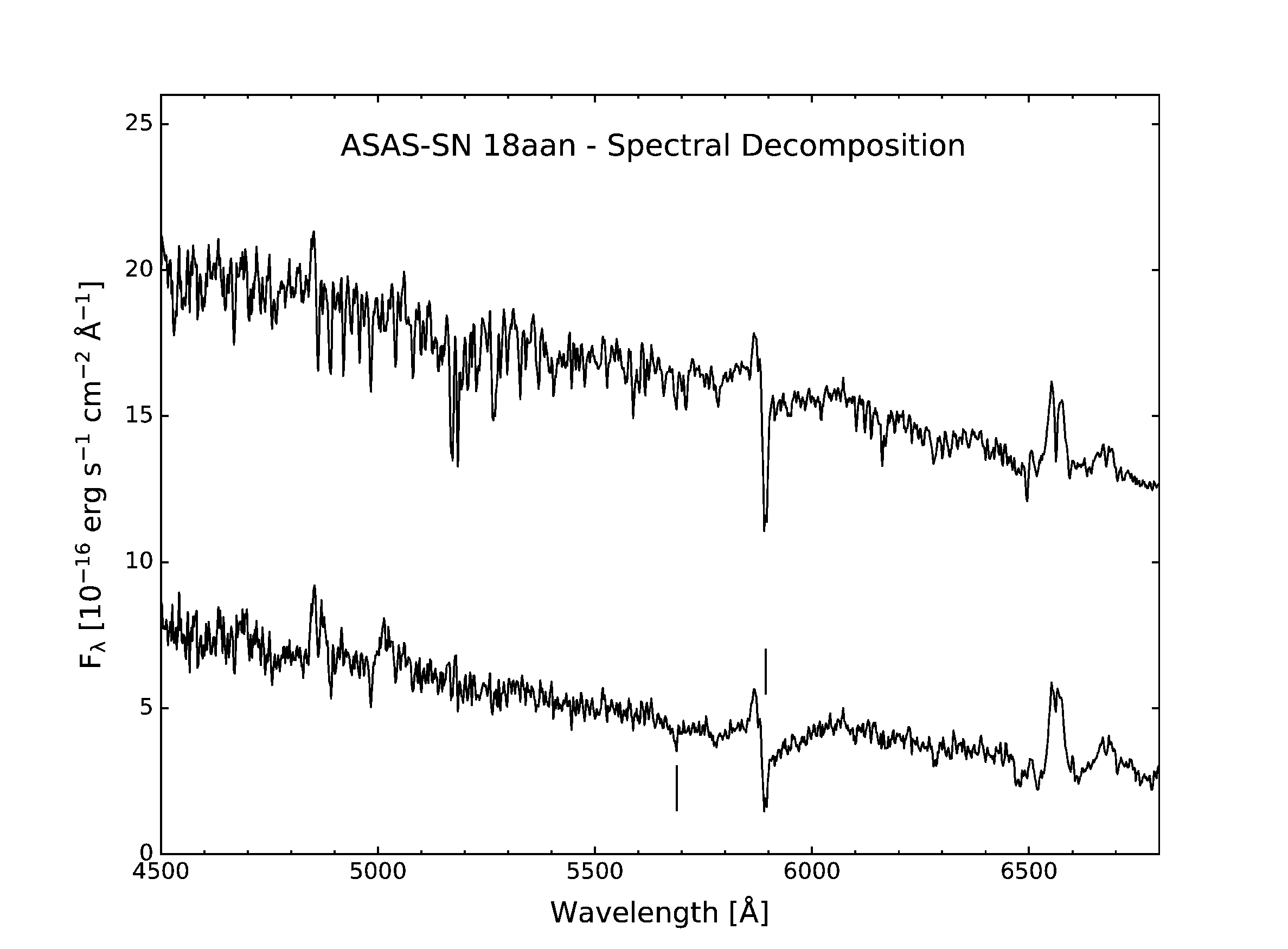}
\end{center}

\caption{(Upper trace): Rest-frame average spectrum, dereddened by 1.19
magnitudes. (Lower trace): The same, after subtraction of a scaled spectrum of
a G9V-type star.  The solid vertical lines mark the wavelengths
of sodium features near $\lambda 5690$ and the very strong
NaD doublet near $\lambda 5893$.
}
\label{fig:specdecomp}
\end{figure}

To study the secondary star's spectrum, we 
averaged our fluxed spectra, after shifting them
to the rest frame of the secondary star using the 
parameters above.  We de-reddened the spectrum
using $E(B-V) = 0.36$ (see the Introduction).
We then used an interactive
program to subtract spectra of different (known)
spectral types and normalizations from this 
rest-frame average, with the aim of cancelling 
the stellar absorption features and leaving the
smooth continuum characteristic of accretion-disk
spectra between the emission lines.  Some of
the template spectra we used were from 
our own observations of stars classified
by \citet{kee89}, and others were originally 
published by \citet{jac84}.  Figure \ref{fig:specdecomp} 
shows the best result, in which the subtracted
spectrum is of the G9V star HD29050 
(from \cite{jac84}) scaled to an {\it unreddened}
flux equivalent to $V = 16.3$, which would be
$V = 17.5$ after reddening.

\section{Analysis}
\subsection{System parameters from superhumps and eclipse modeling}
\label{sec:sys_sh}
We estimated the mass ratio of ASASSN-18aan in the same
way as proposed in \citet{kat13qfromstageA} by using the
fractional superhump-period excess for the 3:1 resonance,
$\varepsilon* = 1 - P_{\rm orb}/P_{\rm stA}$. We calculated
$\varepsilon* = 0.0821(3)$ using the orbital period determined
from the eclipses and the stage A superhump
period. We estimated the mass ratio of ASASSN-18aan
to be $q=0.278(1)$ by using equation (1) and (3) in 
\citet{kat13qfromstageA}.

We studied eclipses during the superoutburst. We performed
intensive, multicolor time-series photometric observations especially at the
timing of eclipses via VSNET and OISTER.

Before analysis of eclipses, we subtracted the global trend of the light curve
in the same way as section \ref{sec:lightcurve}.
We next subtracted the influence of the primary and the bright secondary,
which causes ellipsoidal variations in quiescent state.
We simply subtracted the phase-averaged profile in quiescence, BJD
2458490-2458520, as described in section \ref{sec:sh}.
This is because there seems to be no eclipse of the accretion disk in the 
quiescent state, judged from figure \ref{fig:phave} or \ref{fig:mdmeclipse}.
We see no indication of accretion-disk features such as the hot spot, and
thus the light source of quiescent state is considered to be only components
of the primary and the secondary. After the subtraction of the contribution
of ellipsoidal variations, we also subtracted the variation of superhumps.
We used a same way in principle as the case of subtraction of 
ellipsoidal variations. The superhump profile, however, varies from
moment to moment (see, e.g., figure 8 in \citet{pat95v1159ori}).
We therefore divided the superoutburst into short intervals
and computed phase-averaged superhump profiles in each interval.
We finally removed them from the light curve by dividing the light curve
by them.

The modeling of the eclipsing light curve was constructed mostly in the
same way as \citet{Pdot6} and \citet{kat15ccscl}.
We assumed that radial dependence of the disk surface brightness follows
$r^{p}$, where $r$ is the disk radius and $p$ is the exponent.
Under the assumption of the accretion disk to be the black body 
and steady in time, $p=-3/4$. Since we subtracted the component of
the primary from the light curve in advance, we assumed the disk has an
inner rim.
By using the white dwarf mass estimated in section \ref{sec:vel} and
\ref{sec:ellip}, we estimated the white dwarf radius (equal to the inner
rim of the disk) to be $R_{\rm inner}=0.011a$, where $a$ is a binary
separation \citep{nau72}. The secondary fills its Roche lobe.

We set the disk radius $r$, exponent $p$ and inclination $i$ as free parameters
and estimated them by Markov Chain Monte Carlo (MCMC) method.
We use the mass ratio, $q=0.278(1)$, estimated from the stage A superhump
because the relation between the mass ratio and inclination is quite strong that
it takes much more time to converge.
See Appendix for the detail of the settings of this method.

We first tried to estimate the inclination $i$. At the maxima or minima
of superhumps it is difficult to make better phase-averaged profile from
a few neighboring, continuously changing superhumps, leading to
incorrect estimation. We thus selected three excellent eclipses which
exist on the {\it hillside} of superhumps and have almost linear wings,
using eclipse 9a, 9b and 52 in table E1.
We estimated the disk radius, exponent and inclination simultaneously,
and derived the value of inclination to be $76.54^{+0.06}_{-0.05} \degree,
76.70^{+0.03}_{-0.03} \degree, 76.56^{+0.04}_{-0.04} \degree$ (errors are
calculated from the 95\% credible intervals), and set the inclination
to be $76.60(9) \degree$. Hereafter, in this section we use this value as
the inclination and fixed in our estimation.

\begin{figure}
\begin{center}
\FigureFile(80mm,70mm){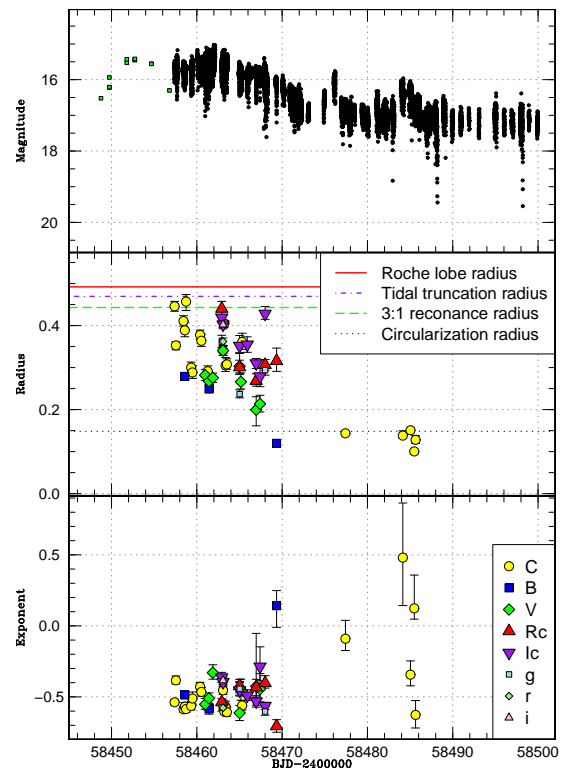}
\end{center}
\caption{Upper panel: Light curve. The filled-square points show
the ASAS-SN's V magnitude. Middle panels: Radius in units of binary separation.
Bottom panel: Exponent. All error bars are calculated from 95 \% confidence
intervals of the posterior probability. The horizontal axis in units of BJD is
common to all of these panels. The points labeled ``Ic'' include
data from both the I and Ic passbands.}
\label{fig:diskvary}
\end{figure}

Figure \ref{fig:diskvary} shows the result of our estimation.
We analyzed 48 eclipses through the superoutburst and two rebrightenings.
We did not investigated the eclipse in quiescence because of its 
shallow depth and low signal-to-noise ratio.
We additionally note that during the rebrightenings we obtained
limited-quality data on only five eclipses.

The effective radius of the Roche lobe \citep{egg83}, the tidal truncation
radius $R_{\rm tidal}/a=0.60/(1+q)$ (approximated equation derived from
\citet{pac77ADmodel}), the 3:1 resonance radius
and the circularization radius \citep{hes90gd552} are indicated in 
Figure \ref{fig:diskvary}.
The disk radius seems to shrink through the superoutburst.
Here note that the disk radius mainly corresponds to the width of the ingress
and egress of the eclipse, and thus the effect of subtracting the
ellipsoidal variations and superhumps is rather small, implying the
estimated values are fairly reliable.
On BJD 2458457, 5 days after the onset of the superoutburst inferred from
the ASAS-SN observation, the disk radius is around $0.4a$, decreasing almost
linearly and then reaching around $0.2a$. The radius at the first and second
rebrightening is near the circularization radius.
A tendency with difference of color are clearly seen; eclipses
observed in the B-band filter correspond to small radii, those in Ic or Rc-band are wider
corresponding to large radii, and ones by C or V-band filter are intermediate
between these two. This results are good agreement with the basic
picture of the accretion disk, which should be hottest in the inner
part and cooler at larger radii.

The disk radius is largest at the onset of the superoutburst,
and comparable to the 3:1 resonance radius.
To investigate this, we performed a linear regression analysis to the data
only including eclipses by C and V-band filter in main superoutburst, and
estimated the disk radius at the onset of the superoutburst to be $0.47a$.
This value is larger than the 3:1 resonance radius, $0.44a$, indicating that the
superoutburst of ASASSN-18aan may be caused by indeed the tidal instability.

The power-law brightness exponent $p$ strongly affects the eclipse depth.
Our estimated values seem to be concentrated around $-0.5$, being small
compared with the value assuming the steady-state black body, $-0.75$.
This might be an artifact of subtracting the ellipsoidal variations and superhumps,
which may artificially reduce the eclipse depth.
In addition, the MDM observations show shallow eclipses
in quiescence at the phase 0.96-1.04 (set phase 1.0 to the mid eclipse).
However, the primary eclipse with the radius of $R_{\rm WD}/a=0.011$ 
occurs at the phase 0.98-01.02, indicating the eclipses in quiescence
include somewhat disk components.
Our estimated values therefore might be systematically small because
of oversubtraction.

\subsection{System parameters from velocity amplitude}
\label{sec:vel}
\begin{figure}
\begin{center}
\FigureFile(90mm,100mm){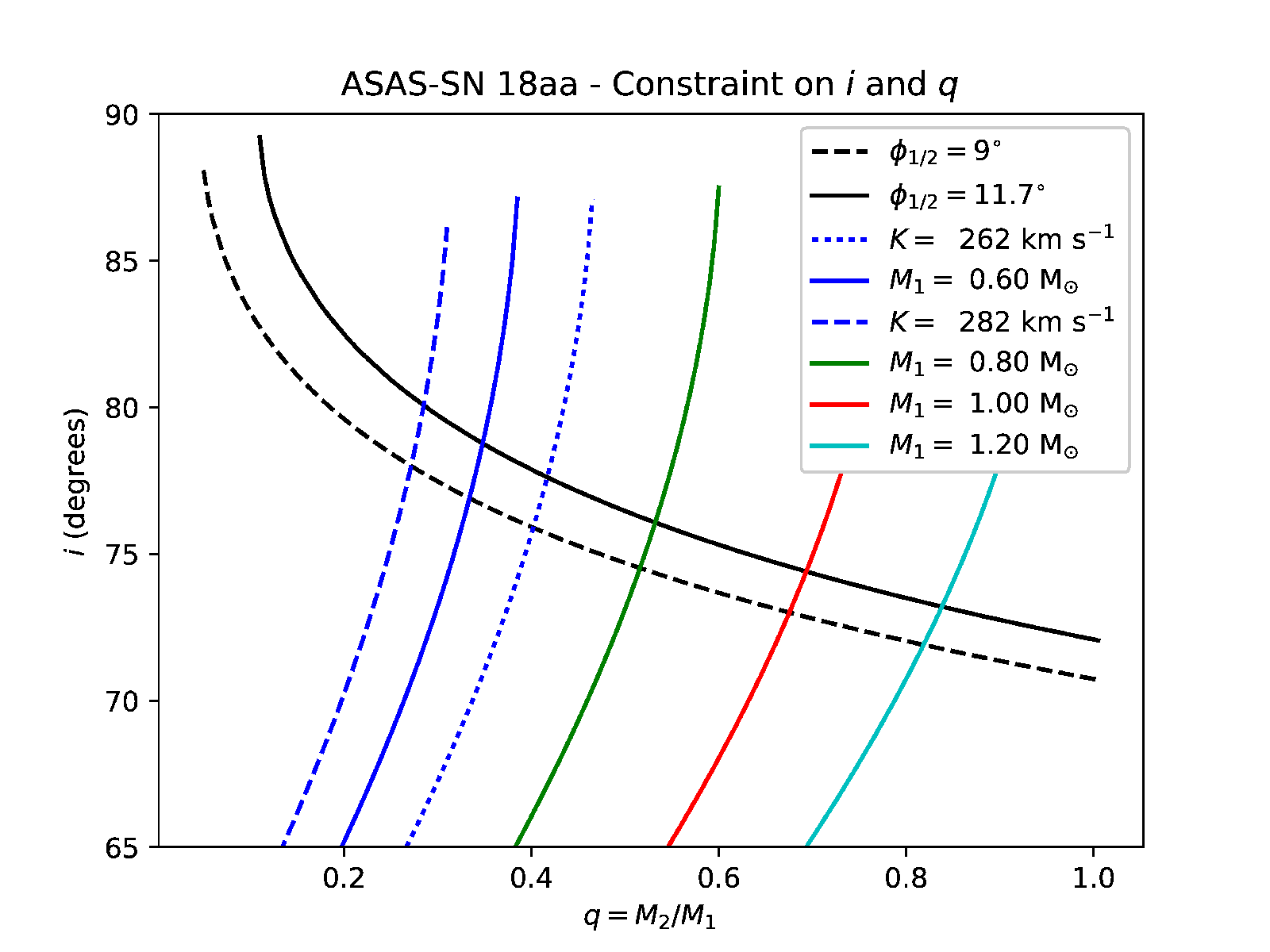}
\end{center}
\caption{Constraints from $K_2$ (vertical curves)
and the eclipse width (downward-sweeping curves) in
the $i$ -- $q$ plane.} 
\label{fig:qvsi}
\end{figure}

We further constrain the system parameters using our 
measured velocity amplitude for the secondary star, $K_2$. 
Figure \ref{fig:qvsi} shows the constraints in $i$-$q$ plane.
The solid, roughly vertical curves are for various assumed
white dwarf masses, $M_{\rm WD}$, for our best fit value of
$K_2$.  The dashed and dotted lines flanking the leftmost 
(blue) curve illustrate the effect of a 
generous 10 km s$^{-1}$ uncertainty in $K_2$.  The curves
sweeping down from the upper left are the constraints
for different eclipse half-widths, $\Delta \phi_{1/2}$. 
The constraint on $q$ implies $M_{\rm WD} \sim 0.6$ 
M$_{\odot}$, so that $M_2 = q M_{\rm WD} \sim 0.18$ 
M$_{\odot}$.  The white dwarf mass is similar to 
average field white dwarfs, and on the low side compared
to most CVs, for which \citet{zor11SDSSCVWDmass} 
give an average $M_{\rm WD} = 0.83 \pm 0.23\ {\rm M_{\odot}}$,
where the standard deviation is for the distribution rather
than the standard deviation of the mean.  The secondary
mass is {\it much} less than that of a main-sequence
G9 star; a main-sequence star of such low mass would
have a spectral type around M4 or M5 \citep{pec13}. 

\subsection{Modeling ellipsoidal variations}
\label{sec:ellip}
\begin{figure}
\begin{center}
\FigureFile(90mm,100mm){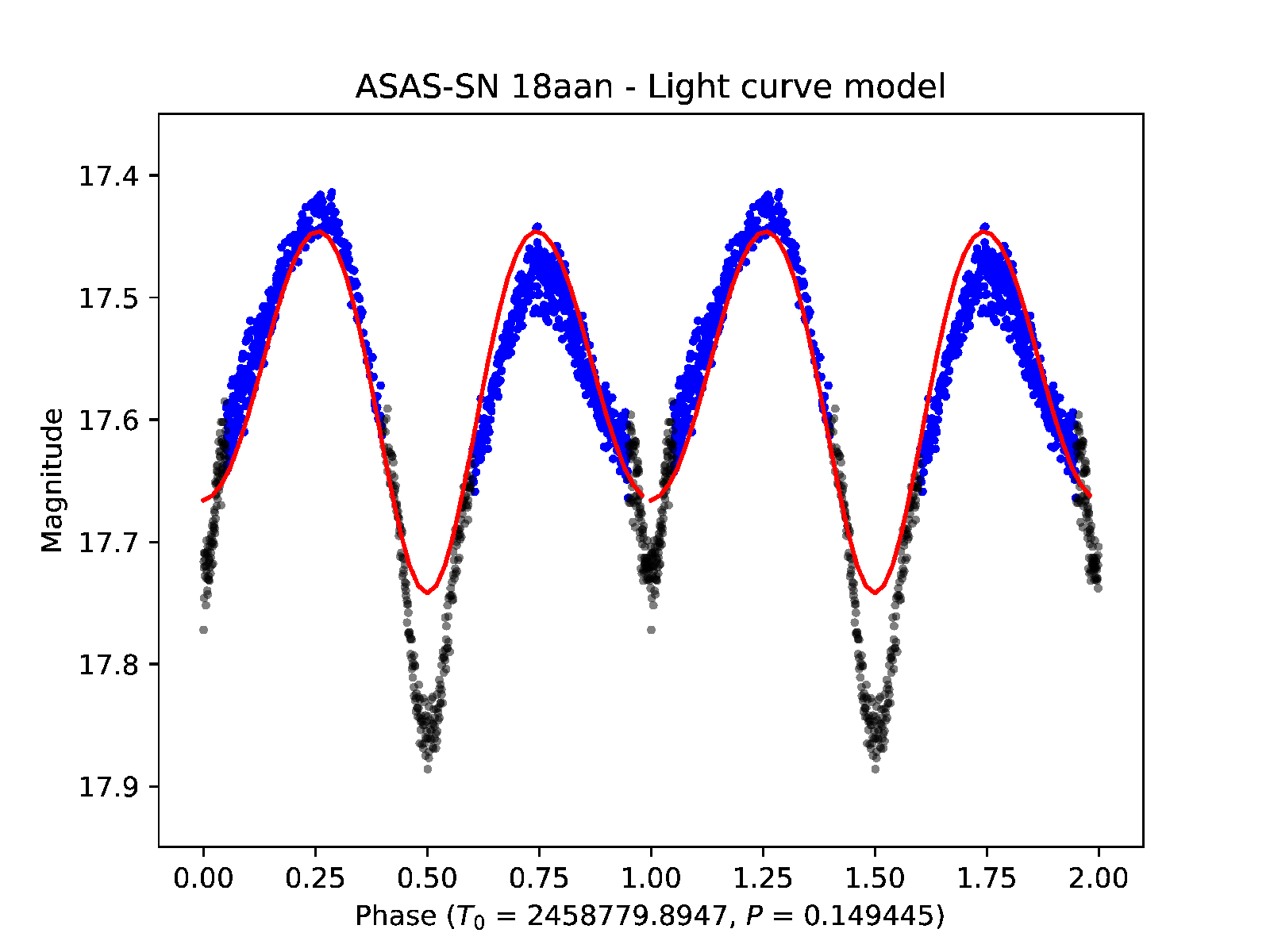}
\end{center}
\caption{Light curve from 2019 October.  The red curve
shows the model of the ellipsoidal variation described in the
text.  The larger blue points were used in the normalization,
and the smaller black points near phase 0.5 and phase 0 were 
ignored.  The modeled light curve has been extinguished by 
$A_V = 1.19$ mag to match the observed points.
} 
\label{fig:modelplot}
\end{figure}

We built a model of the ellipsoidal variations using a 
code developed by \citet{tho05}, which models only the 
contribution from the Roche-lobe-filling secondary star. 
The code computes the Roche lobe's physical size and 
shape for assumed primary and secondary masses, and the
surface brightness of each element as viewed from 
earth including gravity and limb darkening, using
empirical surface brightnesses appropriate to the 
spectral band and surface temperature.  After computing
the brightness at each phase, the code adjusts the 
normalization by finding the distance for which the 
predicted light curve matches the data.
We used the 2019 October MDM light curve, which has the
smallest disk contribution and excellent night-to-night
reproducibility, scaled the broad-band points to approximate
$V$ magnitudes as described above, and subtracted 1.19
mag to correct for extinction.  Points that appeared to 
be affected by the primary or secondary eclipses were
masked.  We set the secondary's $T_{\rm eff}$ to 5300 kelvin,
appropriate for its G9 type \citep{pec13}, as well
as $M_{\rm WD} = 0.6$ and $M_2 = 0.18$ M$_{\odot}$ and
$i = 77^{\circ}$.  To the light curve we added a non-variable
cwomponent equivalent to $F_{\lambda} = 5 \times 10^{-16}$ 
erg cm$^{-2}$ s$^{-1}$ \AA$^{-1}$, about the same as
the disk contribution shown in Figure \ref{fig:specdecomp}, 

Figure \ref{fig:modelplot} shows the result.  We found we could
not match the deep minima around phase 0.5 without a secondary
eclipse.  Also, without the extra light from the disk, the 
predicted amplitude of the ellipsoidal variation at the
system's inclination was too great.  We have one other
bit of evidence that supports the secondary-eclipse
interpretation: referring back to figure \ref{fig:singletrail},
we notice transient {\it absorption} reversals near
phase 0.5 in \ion{He}{1} $\lambda\lambda$ 5876 and 6678
and also enhanced H$\alpha$ absorption. This strongly
suggests that the secondary's light is passing through a
substantial column of high-excitation gas at those phases,
as might be expected in the atmosphere of the disk. 

The distance obtained by scaling the ellipsoidal variation
model to the observed light curve is 624 pc, which, in view
of the long chain of reasoning that went into the choice
of parameters, agrees 
well with the Gaia DR2 parallax determination (675 +32, $-$29 
pc).  The overall
picture is remarkably self-consistent -- $M_{\rm WD} \sim 0.6$ and
$M_2 \sim 0.18 M_{\odot}$, a secondary with $T_{\rm eff}$ 
above $\sim 5000$ kelvin, and $i \sim 77^{\circ}$.

\section{Discussion}
\subsection{3:1 resonance condition in long-period objects}
\label{sec:res}
As mentioned in section \ref{sec:sys_sh}, the disk radius at the onset of the 
superoutburst seems to reach the 3:1 resonance radius. This indicates
the superoutburst of ASASSN-18aan is indeed caused by the tidal
instability in the accretion disk, and thus the upper limit of critical mass
ratios of the 3:1 resonance would be larger than $q=0.278$.
This result presents one possibility that superoutbursts of other
long-$P_{\rm orb}$ objects also may be caused by the tidal instability at
the 3:1 resonance radius.
We therefore investigated SU UMa-type DNe that have long orbital periods
and the mass ratios are (even partially) estimated. We summarized the
results in table \ref{tab:longP}. There are some objects beyond the upper
limit of the tidal instability. Most of them, however, are less reliable values.
First of all, in almost half of the listed objects, mass ratios are estimated
with specific ranges. Objects except TU Men and NY Ser have lower mass
ratio than the upper limit of 0.25 \citep{whi88tidal} or 0.33 with
reduction of mass transfer \citep{mur00SHintermediateq}.

\citet{kat19nyser} estimated a referential value of 
mass ratio of NY Ser from the superhump period of the post-superoutburst
stage as proposed by \citet{kat13qfromstageA}. In this method, the mass
ratio is calculated by assuming ignorance of the pressure effect in the
accretion disk, which conditions are considered to be fulfilled during the
stage A and post-superoutburst stage, and using the orbital period,
superhump period and disk radius. The range of the estimated mass ratio
of NY Ser arose from the supposed disk radius. The disk radius at the
post-superoutburst, however, is uncertain \citep{kat13qfromstageA}, and
thus the mass ratio is quite indistinct.

TU Men has seriously large mass ratios for the tidal instability.
\citet{sto84tumen} derived the mass ratio to be $q=0.59$ from
spectroscopic observations during the 1980 superoutburst.
\citet{men95tumen} estimated the mass ratio to be
$q=0.455(45)$ from H$\alpha$ emission line in quiescent state. He
also reported second candidate of its mass ratio to be $q=0.33$.
\citet{sma06tumen} reanalyzed the radial velocity measured by
\citet{men95tumen} and calculated the mass ratio to be $q=0.5(2)$,
or the lower limit to be $q>0.41(8)$ independently. The mechanism of
superoutburst in TU Men is yet unclear. More precise observations
should be needed to conclude whether the tidal instability acts on the
accretion disk in TU Men.

We should note in the case of U Gem. This object showed long outburst
in 1985, and \citet{sma04ugemSH} reported detection of superhump signal
in this outburst. The mass ratio of U Gem is 0.364(17) \citep{sma01ugem},
beyond the upper limit of the tidal instability in a disk. If U Gem truly showed
superhumps in past outburst, other mechanism like enhanced mass transfer model
should be needed. It is, however, proposed that this signal is doubtful
(\cite{sch07ugemSH}, \cite{Pdot5}), and there are much more rooms for
further discussion before conclusion.

From above confirmation, we conclude that the tidal instability may act
well in long-$P_{\rm orb}$ SU UMa-type DNe, excepting TU Men.
\citet{kat17j0026} indicated that above the period gap, DNe on the standard CV
evolutionary track hardly experience the 3:1 resonance; long-$P_{\rm orb}$ SU UMa-type
DNe seem not to
be on the standard track. As discussed in section \ref{sec:warmsec},
ASASSN-18aan has anomalously warm secondary, which evidently puts it on an
unusual evolutionary track. This is also the case in ASASSN-14ho, another
long-$P_{\rm orb}$ SU UMa-type DNe having a warm secondary
\citep{Gas19}. These results coincide with the argument by \citet{kat17j0026}.

\begin{table*}
\caption{List of long $P_{\rm orb}$ SU UMa-type DNe candidates which
the mass ratios are estimated.}
\begin{tabular}{p{7.8cm}p{1.6cm}p{1.6cm}p{2.6cm}p{1.6cm}}
\hline \hline
Object & $P_{\rm orb}$\commenta & $P_{\rm stB}$\commenta & $q$ & References \\
\hline
SDSS J162520.29+120308.7 & 0.09113(30) & 0.09604(3) & 0.21(1) &1 \\
NY Ser & 0.09744(5) & 0.10458 & 0.23-0.43 & 2, 3 \\
CzeV404 & 0.0980203(6) & 0.10472(2)\commentb & 0.32 & 4 \\
V1006 Cyg & 0.09832(15) & 0.10541(4) & 0.26-0.33 & 5, 6 \\
MN Dra & 0.0998(2) & 0.105040 & 0.258(5)-0.327(5) & 7, 8 \\
TU Men & 0.1172(2) & 0.12558(4)\commentb & 0.5(2) & 9, 10 \\
OT J002656.6+284933 (= CSS101212:002657+284933) & - & 0.13225(1)\commentb & 0.10-0.15 & 11 \\
ASASSN-14ho & 0.24315(10) & - & 0.28 & 12 \\
ASASSN-18aan & 0.149452(2) & 0.15790(5) & 0.278(1) & This paper. \\
\hline
\multicolumn{5}{l}{\commenta In units of day.}\\
\multicolumn{5}{l}{\commentb Averaged period.} \\
\multicolumn{5}{l}{1.\citet{mon17j1625}; 2. \citet{kat19nyser}; 3. \citet{pav14nyser}; 4. \citet{bak14czev404}; 5. \citet{kat16v1006cyg}} \\
\multicolumn{5}{l}{6. \citet{pav18v1006cyg}; 7. \citet{pav10mndra}; 8. \citet{Pdot6}; 9. \citet{men95tumen}; 10. \citet{sma06tumen}} \\
\multicolumn{5}{l}{11. \citet{kat17j0026} ; 12 \citet{Gas19}}
\end{tabular}
\label{tab:longP}
\end{table*}

\subsection{Tidal truncation radius in SU UMa-type and WZ Sge-type DNe}
\citet{pac77ADmodel} calculated the restricted three-body problem and
derived the last non-intersecting orbit around the primary assuming
an inviscid, pressure-free disk. This last non-intersecting
orbit is considered to be the largest radius of a disk, known as the tidal
truncation radius. Several authors have considered the tidal influence at the
outer rim of the disk, and obtained nearly identical results (\cite{ich94tidal};
\cite{pap77tidal}; \cite{tru07}).
There is, however, some room for discussion about the tidal truncation
radius, especially in extremely low-$q$ systems.
\citet{lin79lowqdisk} showed that at extremely small mass ratios,
the outer rim is not truncated and extends beyond the primary Roche lobe.
\citet{hel01eruma} suggested the multiple rebrightenings called
{\it echo outbursts} in ER UMa systems cause a weak angular momentum
dissipation in the hot, eccentric disk due to the extremely low mass ratio. 
The cause of the discrepancy between the observation and theoretical
predictions in WZ Sge-type DNe is unknown. The condition that the 3:1
resonance radius lies inside the tidal truncation radius is realized in the
case of $q<0.25$, being consistent with observations, most of
SU UMa-type DNe having lower than this value. On the other hand,
the condition that the 2:1 resonance radius
lies inside the tidal truncation radius holds only if $q<0.025$.
This is much lower than the value derived from observations, which
concentrates around 0.06--0.08 \citep{kat15wzsge}. \citet{osa02wzsgehump}
suggested that in extremely low mass-ratio system such as WZ Sgr-type DNe, the 
tidal torque at the tidal truncation radius may be too weak to stop the sudden
expansion of the disk due to the outburst, and the outer rim of the disk would 
go beyond the tidal truncation radius and reach the 2:1 resonance radius.
\citet{wak17asassn16eg} indicated that tidal dissipation at the tidal truncation
radius in ASASSN-16eg, WZ Sge-type DNe having anomalously large mass
ratio of $q=0.166(2)$, hardly worked. The 2:1 resonance radius of
ASASSN-16eg lies outside the radius of the primary Roche lobe and,
of course, the tidal truncation radius.

By combining our results with these, we indicate that the tidal torque at the
tidal truncation radius indeed acts well in SU UMa-type DNe; however,
it does not operates effectively in WZ Sge-type DNe. As discussed in
section \ref{sec:res}, even in long-$P_{\rm orb}$ SU UMa-type DNe
the mass ratios seem to be under or near the upper limit, and thus the 3:1
resonance radius likely lies in the tidal truncation radius. This means the
tidal dissipation at the tidal truncation radius acts effectively and the
expansion of the disk limited by this radius. However, when the mass
ratio becomes small and the tidal torque by the secondary also weakens,
the 2:1 resonance radius is the last position that the angular momentum of
the disk is effectively extracted.
\citet{pac77ADmodel} showed that the range of unstable orbits, which first
appear around $q=0.22$, shrinks in extremely low-$q$ systems. This results
may indicate that this unstable orbits are unlikely to work as the point of tidal
limitation for extremely low-$q$ systems such as WZ Sge-type DNe. If so, it is a
reasonable probability that the tidal truncation radius in such extreme systems
is no longer a hard limit to terminate disk expansion.

\subsection{Similarity to WZ Sge-type dwarf novae}
\label{sec:sim}
We regarded the duration of stage A as BJD 2458455.0--2458462.5 (22 cycles)
from the long, uniform superhump period and the amplitude. Considering
a delay of start of our observations, the duration of stage A is somewhat
longer. Judged from the ASAS-SN data, the duration of stage A superhumps
is at most about 10 days. The growth time of superhumps is rather
close to the case of WZ Sge-type DNe \citep{kat15wzsge}.

Rebrightenings after the main superoutburst are also the intrinsic feature of
WZ Sge-type DNe. The rebrightenings of ASASSN-18aan are similar to 
type-B rebrightening, which shows two or more repetitive short brightenings
after a rapid decline of main superoutburst (\cite{kat15wzsge};
\cite{kim16asassn15jd}).

Most recently, the similarity of superoutbursts in long-$P_{\rm orb}$ SU UMa-type
DNe to those in WZ Sge-type one was found in several objects
(\cite{ant02var73dra}; \cite{mro13OGLEDN2}; \cite{kat16v1006cyg};
\cite{kat17j0026}; \cite{kat19csind}; \cite{kat19nyser}; \cite{kat20asas14ho})
\citet{kat16v1006cyg} proposed that such a behavior like WZ Sge-type
superoutbursts is responsible for a difficulty in maintaining the tidal instability
due to the weakness of tidal torque at the 3:1 resonance radius which lies close
to the tidal truncation radius. \citet{kat19csind} also found that the growth
of the 3:1 resonance in systems with the stability border of the 3:1 resonance
seems to be slow.

In the case of ASASN-18aan, the characteristics of the superoutburst are
consistent with other long-$P_{\rm orb}$ SU UMa-type DNe. The similarity
to the WZ Sge-type DNe in such a long-$P_{\rm orb}$ SU UMa-type DNe
would be a key for an investigation of the effect at the tidal truncation radius.

\begin{table*}
\caption{List of CVs with anomalously warm secondary.}
\begin{tabular}{p{4.8cm}p{1.8cm}p{0.7cm}p{0.7cm}p{1.8cm}p{2.6cm}p{1.6cm}}
\hline \hline
Object & $P_{\rm orb}$\commenta & $M_1$\commentb & $M_2$\commentb & Spectral type & Spectral type predicted from $P_{\rm orb}$\commentc & References \\
\hline
EI Psc (=1RXS J232953.9+062814) & 0.044567(3) & 0.7 & 0.13 & K4$\pm$2 & - & 1 \\
QZ Ser & 0.0831612(11) & 0.7 & 0.125 &K4$\pm$2 & M4-M5 & 2 \\
SDSS J170213.26+322954.1 & 0.10008209(9) & 0.94 & 0.20 & M0 & M4 & 3 \\
CSS J134052.0+151341 & 0.10213(8) & 0.7 & 0.5 & K4$\pm$2 & M3-M4 & 4 \\
ASASSN-13cl & 0.20219(8) & - & - & K4$\pm$2 & M3 & 5 \\
ASASSN-14ho & 0.24315(10) & 1.00 & 0.28 & K4$\pm$2 &  M3 & 6 \\
ASASSN-18aan & 0.149452(2) & 0.6 & 0.18 & G9 & M5 & This paper. \\
\hline
\multicolumn{7}{l}{\commenta In units of day.} \\
\multicolumn{7}{l}{\commentb In units of $M_{\odot}$.} \\
\multicolumn{7}{l}{\commentc \citet{kni06CVsecondary}.} \\
\multicolumn{7}{l}{1. \citet{tho02j2329}; 2. \citet{tho02qzser}; 3. \citet{lit06j1702}; 4. \citet{tho13j1340};} \\
\multicolumn{7}{l}{5. \citet{tho15asassn13cl}; 6. \citet{Gas19}}
\end{tabular}
\label{tab:warmdonor}
\end{table*}

\subsection{Anomalously warm secondary}
\label{sec:warmsec}
A 0.18 M$_{\odot}$ main-sequence star will have a 
late M spectral type, dramatically cooler than the
secondary spectrum we observe here.  Mass loss
evidently causes CV secondaries
to depart from the main sequence mass-radius-temperature
relation \citep{kni11CVdonor}, but the 
empirical donor sequence of \citet{kni06CVsecondary}
shows CV secondaries near $P_{\rm orb} = 3.6$ hr
are typically near type M5, with a scatter of a few
subclasses.  The secondary is {\it much}
warmer than expected, for both its determined mass
and the orbital period.
ASASSN-18aan joins a small group of CVs known 
to have anomalously warm secondary stars. We
summarized them in table \ref{tab:warmdonor}.

An apparent enhancement of the sodium absorption line can be seen in
the averaged spectrum after subtraction of a scaled spectrum of G9V-type
star in figure \ref{fig:specdecomp}. Enhanced sodium absorption is 
also seen in other warmer-than-expected systems, such as QZ Ser \citep{tho02qzser}
and CSS J134052.0+151341 \citep{tho13j1340}. \citet{tho02qzser}
suggested that this indicates a history of significant nuclear evolution
of the secondary before mass transfer occurred; 
a side chain of hydrogen burning via CNO cycle, 
${}^{23}{\rm Ne}(p,\gamma){}^{23}{\rm Na}$, leads to enhanced 
sodium.  CV secondary stars in general depart somewhat
from main-sequence mass-radius-luminosity relations, but the
departures in these warm-secondary systems are much larger.  Nuclear evolution
prior to mass transfer can account for this for these gross
discrepancies, as well, since the strong admixture of fused
helium leads to a mean molecular weight very different from 
the main sequence.  Evolutionary scenarios for these systems
are explored briefly by \citet{tho02j2329}.

\section{Summary of Results}

We report photometry and spectroscopy of the eclipsing 
SU UMa-type DN ASASSN-18aan in its 2018 outburst and 
following quiescence. The superoutburst showed a 
slow growth rate of superhumps and rebrightenings, 
characteristics similar to that of WZ Sge-type DNe.
We derived the orbital period to be $P_{\rm orb}=0.149454(3)$d and
confirmed this object has an extraordiarily long orbital period,
for an SU UMa-type DN, above the 2-3 hr period gap.
ASASSN-18aan is a new member of the small class of
long-$P_{\rm orb}$ SU UMa-type DNe. We also estimated the
mass ratio using three different methods and derived 
nearly identical values; from the superoutburst we find $q=0.278(1)$. 
This is close to the upper limit for which the 3:1 resonance
can cause the tidal instability 
(\cite{whi88tidal}; \cite{mur00SHintermediateq}).

By studying the eclipses we find that the disk radius at the onset of the
superoutburst is close to or beyond the 3:1 resonance radius, and
thus the superoutburst indeed can be caused by the tidal instability.
This result suggests that other long-$P_{\rm orb}$
SU UMa-type DNe also could be explained by the tidal instability model,
without the need for other hypotheses.
By considering long-$P_{\rm orb}$ WZ Sge-type DNe
\citep{wak17asassn16eg}, we further suggest that the tidal truncation
radius is effective at larger mass ratios such as those found in
SU UMa-type DNe, but in extremely low-$q$ systems it is no longer
viable; in those cases the outbursting disk may reach the 2:1 resonance,
which lies outside the tidal truncation radius.

One of our most striking results is that the secondary 
is much warmer than typical for CVs of this orbital period.
Its spectral type is G9, while typical secondary stars at
$P_{\rm orb} = 3.6$ hr are in the mid-M range. 
We also find indications of enhanced sodium abundance.
Both these clearly indicate an unusual evolutionary history.

\section{Concluding remarks.}

ASASSN-18aan is an extremely unusual CV.  The vast
majority of SU UMa-type DNe have $P_{\rm orb}$ less than about
two hours, while the period we find is
nearly twice that long.  The period is derived
from eclipses, and corroborated by a large modulation
in the secondary star's radial velocities, so it is
unquestionably orbital, and not due to some other phenomenon.
In addition, the secondary star
is highly anomalous; it is much hotter than typical 
CV secondaries in this period range, and is furthermore
much less massive than its surface temperature would suggest.
These gross peculiarities are very likely the consequence of
mass transfer that began after the secondary star had undergone
significant nuclear evolution; we are seeing, in effect, the stripped
core of a much larger star.  ASASSN-18aan joins a small
number of similar systems already known.

Decades of careful study of superhumping systems have 
yielded a rich empirical picture of the superhump phenomena,
which has spawned a correspondingly detailed body of
theory.  An anomalous sytem such as this one presents an opportunity for 
critical tests of such theories in relatively unexplored
regions of parameter space.

\section*{Acknowledgement}
This work was supported by the Optical and Near-infrared Astronomy
Inter-University Cooperation Program and the Grants-in-Aid of the Ministry of
Education.
This research made use of the AAVSO 
Photometric All-Sky Survey (APASS), funded by the Robert Martin 
Ayers Sciences Fund and NSF AST-1412587.
JRT thanks the MDM staff for observing support, and the 
Tohono O'odham Nation for allowing their mountain to be used
for research into the sky that surrounds us all.
S.S. was supported by the Slovak Research and Development
Agency under the contract No. APVV-15-0458, by the
Slovak Academy of Sciences grant VEGA No. 2/0008/17 and was
partially supported by the Program of Development of
M. V. Lomonosov Moscow State University ``Leading Scientific Schools",
project ``Physics of Stars, Relativistic Objects and Galaxies".
We are grateful to the All-Sky Automated Survey for Supernovae (ASAS-SN)
project for detecting the superoutburst of ASASSN-18aan.
YW is grateful to H. Kimura at Higashi-Hiroshima Observatory for collaboration on observation.
YW deeply thanks the referee Dr. Elena P. Pavlenko for her quick review for
my doctoral dissertation.

\section{Supporting Information}
Supplementary figure E1 and tables E1-E3 are reported in the online version.

\section*{Appendix}
Our developed modeling method of eclipses during outbursts uses
Metropolis algorithm \citep{Metropolis}, which is a kind of MCMC
(for detail, see, e.g., \citet{Sha17MCMC}).

We assumed observed magnitudes following a normal distribution
with specific errors. The likelihood function is thus as follows:
\begin{equation}
	L = \prod^n_{i=1}\frac{1}{\sqrt{2\pi\sigma_i^2}}
		{\rm exp}{\left[ -\frac{(x_i-y_i)^2}{2\sigma_i^2} \right]},
\end{equation}
where $n$, $x_i$, $y_i$ and $\sigma_i$ are a number of data,
observed flux, flux of the synthesized light curve and the photometric
error at $i$-th data of the eclipse, respectively. We practically used
a log-likelihood function. We set the map size at $201 \times 201$.
The reason why we used this mesh value is that, as described in
section \ref{sec:sys_sh}, in this case, the occultation by the
secondary grazes the center part of the accretion disk,
which is a main part of the light source, and thus the synthesized
light curve especially around the mid eclipse are strongly affected by
the ``coarseness" of the map. The smallest mesh value being able to
ignore this effect is found to be 201. We set the photometric error at
$\sigma_i=0.01$ in each $i$-th data.

New candidates of the parameters are independently generated from
a normal distribution, i.e., the proposal distribution of each
parameter is a normal distribution.
We used non-informative prior distribution.
Although there are some restrictions on each parameter (e.g.,
the inclination are estimated from the spectroscopic analysis),
each parameter seemed to converge well.
A width of step in each proposal distribution are decided to converge
the acceptance rate to be about 0.3 \citep{gel96}.

We set the total number of iterations to be $2 \times 10^4$ and ignore
the burn-in period to be $4 \times 10^3$.
We calculated 95 \% credible intervals from the MCMC samples
picked up every 50 iterations (i.e. thinning is 50) to avoid autocorrelations
between iterations.
We thus derived 320 samples as the random samples from the posterior
distribution, which is what we want to estimate.


\begin{thebibliography}{}

\bibitem[Antipin, Pavlenko(2002)]{ant02var73dra}
  Antipin, S.~V., \& Pavlenko, E.~P.\ 2002, A\&A, 391, 565

\bibitem[{B{\c a}kowska} et~al.(2014)]{bak14czev404}
  {B{\c a}kowska}, K., {Olech}, A., {Pospieszy{\'n}ski}, R., {Martinelli}, F.,
  \& {Marciniak}, A.\ 2014, Acta\ Astron., 64, 337

\bibitem[{Cleveland}(1979)]{LOWESS}
  {Cleveland}, W.~S.\ 1979, J. Amer. Statist. Assoc., 74, 829

\bibitem[{Davis} et~al.(2015)]{dav15ASASSNCVAAS}
  {Davis}, A.~B., {Shappee}, B.~J., {Archer Shappee}, B., \& {ASAS-SN}\ 2015,
  American\ Astron.\ Soc.\ Meeting\ Abstracts, 225, \#344.02

\bibitem[{Eggleton}(1983)]{egg83}
  {Eggleton}, P.~P.\ 1983, \apj, 268, 368

\bibitem[{Gaia Collaboration} et~al.(2018)]{GaiaPaper2}
  {Gaia Collaboration}, {et~al.}\ 2018, \aap, 616, A1

\bibitem[{Gasque} et~al.(2019)]{Gas19}
  {Gasque}, L.~C., {Hening}, C.~A., {Hviding}, R.~E., {Thorstensen}, J.~R.,
  {Paterson}, K., {Breytenbach}, H., {Motsoaledi}, M., \& {Woudt}, P.~A.\ 2019,
  \aj, 158, 156

\bibitem[{Gelman} et~al.(1996)]{gel96}
  {Gelman}, A., {Roberts}, G., \& {Gilks}, W.\ 1996, Bayesian statistics, 5,
  599

\bibitem[{Green} et~al.(2019)]{gre19}
  {Green}, G.~M., {Schlafly}, E., {Zucker}, C., {Speagle}, J.~S., \&
  {Finkbeiner}, D.\ 2019, \apj, 887, 93

\bibitem[{Green} et~al.(2018)]{gre18}
  {Green}, G.~M., {et~al.}\ 2018, \mnras, 478, 651

\bibitem[Hellier(2001)]{hel01eruma}
  Hellier, C.\ 2001, PASP, 113, 469

\bibitem[{Henden} et~al.(2011)]{hen11aavso}
  {Henden}, A.~A., {Levine}, S.~E., {Terrell}, D., {Smith}, T.~C., \& {Welch},
  D.~L.\ 2011, American Astronomical Society Meeting Abstracts \#218, Bulletin
  of the American Astronomical Society, 43, 126.01

\bibitem[Hessman, Hopp(1990)]{hes90gd552}
  Hessman, F.~V., \& Hopp, U.\ 1990, A\&A, 228, 387

\bibitem[{Hirose}, {Osaki}(1990)]{hir90SHexcess}
  {Hirose}, M., \& {Osaki}, Y.\ 1990, PASJ, 42, 135

\bibitem[Ichikawa, Osaki(1994)]{ich94tidal}
  Ichikawa, S., \& Osaki, Y.\ 1994, PASJ, 46, 621

\bibitem[{Jacoby} et~al.(1984)]{jac84}
  {Jacoby}, G.~H., {Hunter}, D.~A., \& {Christian}, C.~A.\ 1984, \apjs, 56, 257

\bibitem[{Kato}(2015)]{kat15wzsge}
  {Kato}, T.\ 2015, PASJ, 67, 108

\bibitem[{Kato}(2020)]{kat20asas14ho}
  {Kato}, T.\ 2020, \pasj, 72, L2

\bibitem[{Kato} et~al.(2014a)]{Pdot6}
  {Kato}, T., {et~al.}\ 2014a, PASJ, 66, 90

\bibitem[{Kato} et~al.(2014b)]{Pdot5}
  {Kato}, T., {et~al.}\ 2014b, PASJ, 66, 30

\bibitem[{Kato} et~al.(2019a)]{kat19csind}
  {Kato}, T., {Hambsch}, F.-J., {Monard}, B., {Nelson}, P., {Stubbings}, R., \&
  {Starr}, P.\ 2019a, \pasj, 71, L4

\bibitem[{Kato} et~al.(2015)]{kat15ccscl}
  {Kato}, T., {Hambsch}, F.-J., {Oksanen}, A., {Starr}, P., \& {Henden}, A.\
  2015, PASJ, 67, 3

\bibitem[{Kato} et~al.(2009)]{Pdot}
  {Kato}, T., {et~al.}\ 2009, PASJ, 61, S395

\bibitem[{Kato}, {Osaki}(2013)]{kat13qfromstageA}
  {Kato}, T., \& {Osaki}, Y.\ 2013, PASJ, 65, 115

\bibitem[{Kato} et~al.(2019b)]{kat19nyser}
  {Kato}, T., {et~al.}\ 2019b, \pasj, 71, L1

\bibitem[{Kato} et~al.(2016)]{kat16v1006cyg}
  {Kato}, T., {et~al.}\ 2016, PASJ, 68, L4

\bibitem[{Kato} et~al.(2017)]{kat17j0026}
  {Kato}, T., {et~al.}\ 2017, \pasj, 69, L4

\bibitem[Kato et~al.(2004)]{VSNET}
  Kato, T., Uemura, M., Ishioka, R., Nogami, D., Kunjaya, C., Baba, H., \&
  Yamaoka, H.\ 2004, PASJ, 56, S1

\bibitem[Katysheva, Pavlenko(2003)]{kat03CVperiodgap}
  Katysheva, N.~A., \& Pavlenko, E.~P.\ 2003, Astrophysics, 46, 114

\bibitem[{Keenan}, {McNeil}(1989)]{kee89}
  {Keenan}, P.~C., \& {McNeil}, R.~C.\ 1989, \apjs, 71, 245

\bibitem[{Kimura} et~al.(2016)]{kim16asassn15jd}
  {Kimura}, M., {et~al.}\ 2016, \pasj, 68, 55

\bibitem[{Knigge}(2006)]{kni06CVsecondary}
  {Knigge}, C.\ 2006, MNRAS, 373, 484

\bibitem[{Knigge} et~al.(2011)]{kni11CVdonor}
  {Knigge}, C., {Baraffe}, I., \& {Patterson}, J.\ 2011, ApJS, 194, 28

\bibitem[{Kochanek} et~al.(2017)]{Koc17ASASSN}
  {Kochanek}, C.~S., {et~al.}\ 2017, \pasp, 129, 104502

\bibitem[{Lin}, {Papaloizou}(1979)]{lin79lowqdisk}
  {Lin}, D.~N.~C., \& {Papaloizou}, J.\ 1979, \mnras, 186, 799

\bibitem[{Littlefair} et~al.(2006)]{lit06j1702}
  {Littlefair}, S.~P., {Dhillon}, V.~S., {Marsh}, T.~R., \& {G{\"a}nsicke},
  B.~T.\ 2006, MNRAS, 371, 1435

\bibitem[{Lubow}(1991)]{lub91SHa}
  {Lubow}, S.~H.\ 1991, ApJ, 381, 259

\bibitem[Marsh et~al.(1990)]{mar90ugem}
  Marsh, T.~R., Horne, K., Schlegel, E.~M., Honeycutt, R.~K., \& Kaitchuck,
  R.~H.\ 1990, ApJ, 364, 637

\bibitem[{Martini} et~al.(2011)]{mar11}
  {Martini}, P., {et~al.}\ 2011, \pasp, 123, 187

\bibitem[Mennickent(1995)]{men95tumen}
  Mennickent, R.~E.\ 1995, A\&A, 294, 126

\bibitem[{Metropolis} et~al.(1953)]{Metropolis}
  {Metropolis}, N., {Rosenbluth}, A.~W., {Rosenbluth}, M.~N., {Teller}, A.~H.,
  \& {Teller}, E.\ 1953, J.\ Chem.\ Phys., 21, 1087

\bibitem[{Montgomery} et~al.(2017)]{mon17j1625}
  {Montgomery}, M.~M., {Voloshina}, I., {Olenick}, R., {Meziere}, K., \&
  {Metlov}, V.\ 2017, New\ Astron., 50, 43

\bibitem[{Mroz} et~al.(2013)]{mro13OGLEDN2}
  {Mroz}, P., {et~al.}\ 2013, Acta\ Astron., 63, 135

\bibitem[{Murray} et~al.(2000)]{mur00SHintermediateq}
  {Murray}, J., {Warner}, B., \& {Wickramasinghe}, D.\ 2000, New\ Astron.\
  Rev., 44, 51

\bibitem[{Nauenberg}(1972)]{nau72}
  {Nauenberg}, M.\ 1972, \apj, 175, 417

\bibitem[{Nesci} et~al.(2019)]{nes19a18aan}
  {Nesci}, R., {Tuvikene}, T., \& {Gualandi}, R.\ 2019, Open European Journal
  on Variable Stars, 196, 1

\bibitem[{Osaki}, {Meyer}(2002)]{osa02wzsgehump}
  {Osaki}, Y., \& {Meyer}, F.\ 2002, A\&A, 383, 574

\bibitem[Paczy\'{n}ski(1977)]{pac77ADmodel}
  Paczy\'{n}ski, B.\ 1977, ApJ, 216, 822

\bibitem[{Papaloizou}, {Pringle}(1977)]{pap77tidal}
  {Papaloizou}, J., \& {Pringle}, J.~E.\ 1977, \mnras, 181, 441

\bibitem[{Patterson} et~al.(1995)]{pat95v1159ori}
  {Patterson}, J., {Jablonski}, F., {Koen}, C., {O'Donoghue}, D., \&
  {Skillman}, D.~R.\ 1995, PASP, 107, 1183

\bibitem[{Pavlenko} et~al.(2014)]{pav14nyser}
  {Pavlenko}, E.~P., {et~al.}\ 2014, PASJ, 66, 111

\bibitem[{Pavlenko} et~al.(2018)]{pav18v1006cyg}
  {Pavlenko}, E.~P., {et~al.}\ 2018, Contributions of the Astronomical
  Observatory Skalnate Pleso, 48, 339

\bibitem[{Pavlenko} et~al.(2010)]{pav10mndra}
  {Pavlenko}, E.~P., {et~al.}\ 2010, Astron.\ Rep., 54, 6

\bibitem[{Pecaut}, {Mamajek}(2013)]{pec13}
  {Pecaut}, M.~J., \& {Mamajek}, E.~E.\ 2013, \apjs, 208, 9

\bibitem[{Schreiber}(2007)]{sch07ugemSH}
  {Schreiber}, M.~R.\ 2007, A\&A, 466, 1025

\bibitem[{Shappee} et~al.(2014)]{ASASSN}
  {Shappee}, B.~J., {et~al.}\ 2014, ApJ, 788, 48

\bibitem[{Sharma}(2017)]{Sha17MCMC}
  {Sharma}, S.\ 2017, Annual Review of Astronomy and Astrophysics, 55, 213

\bibitem[{Smak}, {Waagen}(2004)]{sma04ugemSH}
  {Smak}, J., \& {Waagen}, E.~O.\ 2004, Acta\ Astron., 54, 433

\bibitem[Smak(2001)]{sma01ugem}
  Smak, J.~I.\ 2001, Acta\ Astron., 51, 279

\bibitem[{Smak}(2006)]{sma06tumen}
  {Smak}, J.~I.\ 2006, Acta\ Astron., 56, 277

\bibitem[Stellingwerf(1978)]{PDM}
  Stellingwerf, R.~F.\ 1978, ApJ, 224, 953

\bibitem[{Stolz}, {Schoembs}(1984)]{sto84tumen}
  {Stolz}, B., \& {Schoembs}, R.\ 1984, A\&A, 132, 187

\bibitem[{Thorstensen}(2013)]{tho13j1340}
  {Thorstensen}, J.~R.\ 2013, PASP, 125, 506

\bibitem[{Thorstensen}(2015)]{tho15asassn13cl}
  {Thorstensen}, J.~R.\ 2015, PASP, 127, 351

\bibitem[{Thorstensen} et~al.(2016)]{tho16}
  {Thorstensen}, J.~R., {Alper}, E.~H., \& {Weil}, K.~E.\ 2016, \aj, 152, 226

\bibitem[{Thorstensen}, {Armstrong}(2005)]{tho05}
  {Thorstensen}, J.~R., \& {Armstrong}, E.\ 2005, \aj, 130, 759

\bibitem[Thorstensen et~al.(2002a)]{tho02qzser}
  Thorstensen, J.~R., Fenton, W.~H., Patterson, J.~O., Kemp, J., Halpern, J.,
  \& Baraffe, I.\ 2002a, PASP, 114, 1117

\bibitem[Thorstensen et~al.(2002b)]{tho02j2329}
  Thorstensen, J.~R., Fenton, W.~H., Patterson, J.~O., Kemp, J., Krajci, T., \&
  Baraffe, I.\ 2002b, ApJ, 567, L49

\bibitem[{Tonry}, {Davis}(1979)]{ton79}
  {Tonry}, J., \& {Davis}, M.\ 1979, \aj, 84, 1511

\bibitem[{Truss}(2007)]{tru07}
  {Truss}, M.~R.\ 2007, \mnras, 376, 89

\bibitem[{Wakamatsu} et~al.(2017)]{wak17asassn16eg}
  {Wakamatsu}, Y., {et~al.}\ 2017, \pasj, 69, 89

\bibitem[Whitehurst(1988)]{whi88tidal}
  Whitehurst, R.\ 1988, MNRAS, 232, 35

\bibitem[{Zorotovic} et~al.(2011)]{zor11SDSSCVWDmass}
  {Zorotovic}, M., {Schreiber}, M.~R., \& {G{\"a}nsicke}, B.~T.\ 2011, A\&A,
  536, A42

\end{thebibliography}
\newcommand{\noop}[1]{}

\bibliographystyle{pasjtest1}

\end{document}